\documentclass[10pt,english]{IEEEtran}
\usepackage[LGR,T1]{fontenc}
\usepackage[latin9]{inputenc}
\usepackage{float}
\usepackage{textcomp}
\usepackage{bm}
\usepackage{amsmath}
\usepackage{amssymb}
\usepackage{stackrel}
\usepackage{graphicx}
\usepackage[colorlinks]{hyperref}
\newtheorem{remark}{Remark}
\usepackage[noend]{algpseudocode}
\usepackage{algorithmicx,algorithm}

\makeatletter

\newcommand*\LyXZeroWidthSpace{\hspace{0pt}}
\DeclareRobustCommand{\greektext}{%
  \fontencoding{LGR}\selectfont\def\encodingdefault{LGR}}
\DeclareRobustCommand{\textgreek}[1]{\leavevmode{\greektext #1}}
\DeclareFontEncoding{LGR}{}{}
\DeclareTextSymbol{\~}{LGR}{126}
\floatstyle{ruled}
\newfloat{algorithm}{tbp}{loa}
\providecommand{\algorithmname}{Algorithm}
\floatname{algorithm}{\protect\algorithmname}

\makeatother

\usepackage{babel}
\begin{document}

\title{Secure ISAC with Fluid Antenna Systems:\\Joint Precoding and Port Selection}

\author{Abdelhamid Salem,~\IEEEmembership{Member,~IEEE}, 
            Hao Xu,~\IEEEmembership{Member,~IEEE}, 
            Kai-Kit Wong,~\IEEEmembership{Fellow,~IEEE},\\ 
            Chan-Byoung Chae,~\IEEEmembership{Fellow,~IEEE}, and 
            Yangyang Zhang
\vspace{-6mm}

\thanks{The work of K. K. Wong is supported by the Engineering and Physical Sciences Research Council (EPSRC) under grant EP/W026813/1.}
\thanks{The work of C.-B. Chae is supported by the Institute for Information and Communication Technology Planning and Evaluation (IITP) grants funded by the Ministry of Science and ICT (MSIT), South Korea, under Grant RS-2024-00428780 and RS-2024-00397216.}

\thanks{A. Salem and K. K. Wong are with the Department of Electronic and Electrical Engineering, University College London, WC1E 7JE, London, United Kingdom (e-mail: $\rm\{a.salem,kai\text{-}kit.wong\}@ucl.ac.uk$). K. K. Wong is also affiliated with Yonsei Frontier Laboratory, Yonsei University, Seoul, 03722, Republic of Korea.}
\thanks{H. Xu is with National Mobile Communications Research Laboratory, Southeast University, Nanjing, China (e-mail: $\rm hao.xu@seu.edu.cn$).}
\thanks{C.-B. Chae is with the School of Integrated Technology, Yonsei University, Seoul, 03722, Republic of Korea (e-mail: $\rm cbchae@yonsei.ac.kr$).}
\thanks{Y. Zhang is with Kuang-Chi Science Limited, Hong Kong SAR, China (e-mail: $\rm yangyang.zhang@kuang\text{-}chi.com$).}

\thanks{Corresponding author: Kai-Kit Wong.}
}

\maketitle
\begin{abstract}
This paper presents a novel framework for enhancing physical-layer security in integrated sensing and communication (ISAC) systems by leveraging the reconfigurability of fluid antenna systems (FAS). We propose a joint precoding and port selection (JPPS) strategy that maximizes the sum secrecy rate while simultaneously ensuring reliable radar sensing. The problem is formulated using fractional programming (FP) and solved through an iterative algorithm that integrates FP transformations with successive convex approximation (SCA). To reduce computational complexity, we further develop low-complexity schemes based on zero-forcing (ZF) precoding, combined with greedy port selection and trace-inverse minimization. Simulation results demonstrate substantial improvements in both secrecy performance and sensing accuracy compared to conventional baselines, across a wide range of FAS ports, user loads, and sensing targets. These findings highlight the critical importance of FAS geometry optimization in enabling secure and efficient joint communication-sensing for next-generation wireless networks.
\end{abstract}

\begin{IEEEkeywords}
Fluid antenna system (FAS), integrated sensing and communication (ISAC), physical layer security.
\end{IEEEkeywords}

\section{Introduction}
\subsection{Context}
\IEEEPARstart{F}{luid} antenna systems (FAS) represent a novel paradigm of treating radiating structure or `antenna' as a reconfigurable physical-layer resource, broadening system design and network optimization and inspiring next-generation reconfigurable antennas, with the emphasis on the feature of shape and position reconfigurability in antennas \cite{Wong2020Fluid,Wong2022Bruce,New2025Tutorial}. In \cite{Lu-2025}, Lu {\em et al.}~further provided an interpretation for FAS through the lens of electromagnetics. Originally proposed by Wong {\em et al.}~in \cite{FAS1,Fas2}, FAS has been demonstrated to obtain tremendous spatial diversity even on a single radio-frequency (RF) chain. Practical FASs may be implemented using liquid-based antennas \cite{I24_shen2024design}, reconfigurable pixels \cite{fas8}, and metamaterial structures \cite{Liu-2025arxiv}. In \cite{Tong-2025}, trade-offs among different technologies are discussed.

In recent years, numerous efforts have been made to study the performance of FAS channels. For instance, Khammassi {\em et al.}~adopted an eigenvalue-based channel model to characterize the spatial correlation among FAS ports and investigated the diversity benefits of FAS in Rayleigh fading channels \cite{Khammassi-2023}. Subsequent studies then tried to understand the performance of FAS in Nakagami fading channels \cite{fas5} and even $\alpha$-$\mu$ fading channels \cite{Alvim-2023} utilizing a simplified modelling approach. Exact analysis for FAS is difficult because of the correlation existed between FAS ports. Thus far, it appears to be possible only in the Rayleigh fading case \cite{G5_new2023SISO-FAS}. To overcome this, Ram\'{i}rez-Espinosa {\em et al.}~proposed the spatial block-correlation model that can achieve great accuracy while maintaining tractability \cite{H7_Espinosa2024Anew}. It was also reported in \cite{G10_new2024MIMO-FAS} that when multi-port FAS is employed at both transmitter and receiver ends, extraordinary spatial diversity can be obtained. Furthermore, the continuous version of FAS was considered and studied in \cite{Psomas-dec2023}.

FAS also has led to many meaningful optimization problems enhancing the performance of various wireless communication systems. For instance, \cite{Fas3} employed deep learning to find the optimal port for a receive FAS in a single-user system. Port selection of the fluid multiple-input multiple-output (MIMO) system in which multiple fluid antennas are deployed at both ends, was tackled in \cite{G13_Efrem2024MIMO-FAS,Chen-2025fmimo}. In \cite{Feng-2025}, deep unfolding was used to accelerate the port selection process in time-varying channels. Moreover, FAS in multiuser channels gives rise to problems requiring joint optimization of beamforming and port selection, one of which was addressed in \cite{Xu-fasup2025}. Opportunistic scheduling is also an effective scheme to combine with FAS and was studied in \cite{fas4,Waqar-2024}. FAS has also been shown to be compatible to multi-carrier systems \cite{Hong2025FAS}. Artificial intelligence \cite{Wang-2024oct} and large language models \cite{Wang-2025llm} for FAS are being actively pursued in research. On the other hand, an important use case of FAS is multiple access, which has been explored in \cite{H4_wong2022FAMA,H5_wong2023fast,H6_wong2023sFAMA,H12_Wong2024cuma,H10_hong2024coded,H11_hong2025Downlink,Waqar-tfama2025}. Besides, an emerging direction appears to consider FAS in cell-free networks \cite{Han-2025}. Channel state information (CSI) is required in FAS and CSI estimation for FAS has recently been addressed in \cite{xu2024channel,fas7,fas9}.

Apparently, FAS can be viewed as a new degree of freedom (DoF) that can elevate the performance of wireless communications. In this context, a particularly important application is integrated sensing and communication (ISAC) \cite{Ref1,Ref2}. In ISAC, both radar sensing and wireless communications are combined via shared use of the spectrum, hardware platform and a signal processing framework. Given its great potential, ISAC has now become one of the hottest topics in recent years. In \cite{Generalized}, the authors considered ISAC beamforming design to detect targets as a MIMO radar and communicate to multiple users simultaneously. Later, \cite{joint} introduced a joint transmit beamforming model for a dual-functional MIMO radar and multiple communication users. Then in \cite{Hybrid}, the hybrid transmit/receive beamformers were designed by maximizing the sum-rate in ISAC under transmit power constraints. Moreover, \cite{Spectral} considered the mutual information between the target reflections and the target responses for ISAC systems as the design metric, and obtained the optimal waveforms achieving the maximum mutual information. Further works have looked into ISAC systems based on the orthogonal frequency division multiplexing (OFDM) waveform \cite{OFDM}, developed an integrated scheduling method for sensing, communication, and control \cite{Rev1ref2}, and introduced a new mathematical framework for ISAC cellular networks \cite{meisac}. A survey of the recent progress in the areas of radar-communication coexistence and ISAC systems covering applications and approaches can be found in \cite{survy}. 

However, a major challenge in ISAC networks is ensuring secure communication, as the shared spectrum increases the risk of information leakage to both legitimate sensing targets and malicious eavesdroppers. As a consequence, physical layer security techniques have been proposed to mitigate threats in ISAC systems. For example, \cite{Ref3} investigated secure ISAC via optimized precoding and beamforming strategies. Recently, in \cite{Ref4}, the security of ISAC systems was considered by using phase-coupled intelligent omni-surfaces. Additionally, a joint design of transmit beamforming to minimize the maximum eavesdropping signal-to-interference-plus-noise ratio (SINR), while guaranteeing the communication quality-of-service, the radar detection requirements, as well as the transmit power constraints, was devised in \cite{Ref5}. A novel sensing-aided secure communication protocol was also proposed in \cite{Ref6}.

\subsection{State-of-the-Art for FAS in ISAC}
ISAC is a very attractive proposition but doing so inevitably further drains the limited resources even more. An additional DoF is desperately needed and thus FAS can be an essential technology that can make ISAC truly desirable \cite{Meng-2025}. With that in mind, several researches have been conducted. In \cite{H17_wang2024fluid}, it was revealed that multiuser MIMO with FAS outperforms significantly that without FAS for ISAC. Moreover, FAS has been illustrated to expand the achievability region for ISAC systems \cite{Zhou-2024fisac,Ye-2025,H16_Zou2024shifting}. Furthermore, FAS for ISAC has been considered in full-duplex systems \cite{Tang-2025fisac}, cooperative multi-state systems \cite{Lou-2025}, and backscattering scenarios \cite{H15_ghadi2024perf}. Nevertheless, the use of FAS for secure ISAC is not well understood.

\subsection{Contributions}
In this paper, we propose a novel framework for enhancing the secrecy performance of downlink ISAC systems empowered by FAS where the eavesdropper can be one legitimate user or a target. The aim is to jointly optimize the transmit precoding and the selection of a subset of FAS ports to maximize the sum secrecy rate while satisfying a minimum radar SINR constraint. To provide real insights, we take into account practical fluid antenna spatial correlation using the Jake's model and introduce a scalable methodology to exploit the FAS flexibility. Despite its potential, integrating FAS into ISAC architectures poses significant challenges. First, antenna port selection directly affects both the communication channels and radar beampatterns, leading to highly non-convex optimization problems. Second, the spatial correlation among FAS ports, governed by their relative positions, further complicates the secrecy rate and radar SINR analysis. Moreover, the worst-case secrecy performance must be guaranteed in the presence of internal eavesdroppers (e.g., idle users) and external passive interceptors (e.g., radar targets). To date, these aspects have not been jointly addressed in the literature. The main contributions of this paper are summarized as follows:
\begin{itemize}
\item We propose a novel ISAC framework that leverages the power of FAS to jointly enhance communication secrecy and radar performance in a multiuser environment where the eavesdropper can be one legitimate user or a target. 
\item A joint optimization problem is formulated to maximize the sum secrecy rate of legitimate users under a radar SINR constraint. The problem jointly designs the transmit precoder and selects the optimal subset of FAS ports. To tackle the non-convexity of the formulated problem, we develop an efficient iterative algorithm based on fractional programming (FP) and successive convex approximation (SCA). A novel utility-driven port selection metric is derived from the FP structure to guide the selection of active ports without relying on greedy or heuristic search.
\item To reduce the complexity and provide security across the users, we propose to adopt the zero-forcing (ZF) precoder, while the ports selection scheme is designed to achieve secure transmission, achieving the required sensing SINR. Accordingly, we reformulate the port selection problem and develop a greedy algorithm that offers strong performance with reduced computational cost.
\item For the sake of comparison and to further reduce computational complexity, we reformulate the port selection problem as a trace-inverse minimization, and an efficient greedy method and a novel singular value decomposition (SVD)-based port selection method are proposed. 
\item Through computer simulations, we demonstrate the performance gains in secrecy rate of the proposed schemes. The proposed schemes significantly outperform the conventional fixed-port baseline. The results reveal how fluid antenna directly impacts security-sensing trade-offs. 
\end{itemize}

While prior studies have demonstrated the benefits of FAS in multiuser systems, its potential for securing ISAC systems remains largely unexplored. To the best of our knowledge, this is the first work to investigate physical layer security optimization in ISAC systems using reconfigurable fluid antenna arrays considering that the eavesdropper can be one legitimate user or a target. The proposed methods offer scalable, low-complexity strategies for harnessing the spatial flexibility of FAS under realistic correlation models and security constraints.

The remainder of the paper is organized as follows. Section \ref{sec:System-Model} presents the system and channel models. Section \ref{sec:Problem-Formulation} formulates the secrecy rate maximization problem under radar SINR constraints, and introduces the joint precoding and port selection optimization framework. After that, in Section \ref{sec:Proposed-Solutions}, we devise an alternating optimization algorithms based on FP and SCA, along with the utility-based port selection strategy. Section \ref{sec:Numerical-Results} provides extensive numerical results that validate the proposed methods and highlight its performance advantages over existing baselines. Finally, Section \ref{sec:Conclusion} concludes the paper and outlines directions for future research.

\section{System Model\label{sec:System-Model}}
\begin{figure}
\begin{centering}
\includegraphics[width=.95\columnwidth]{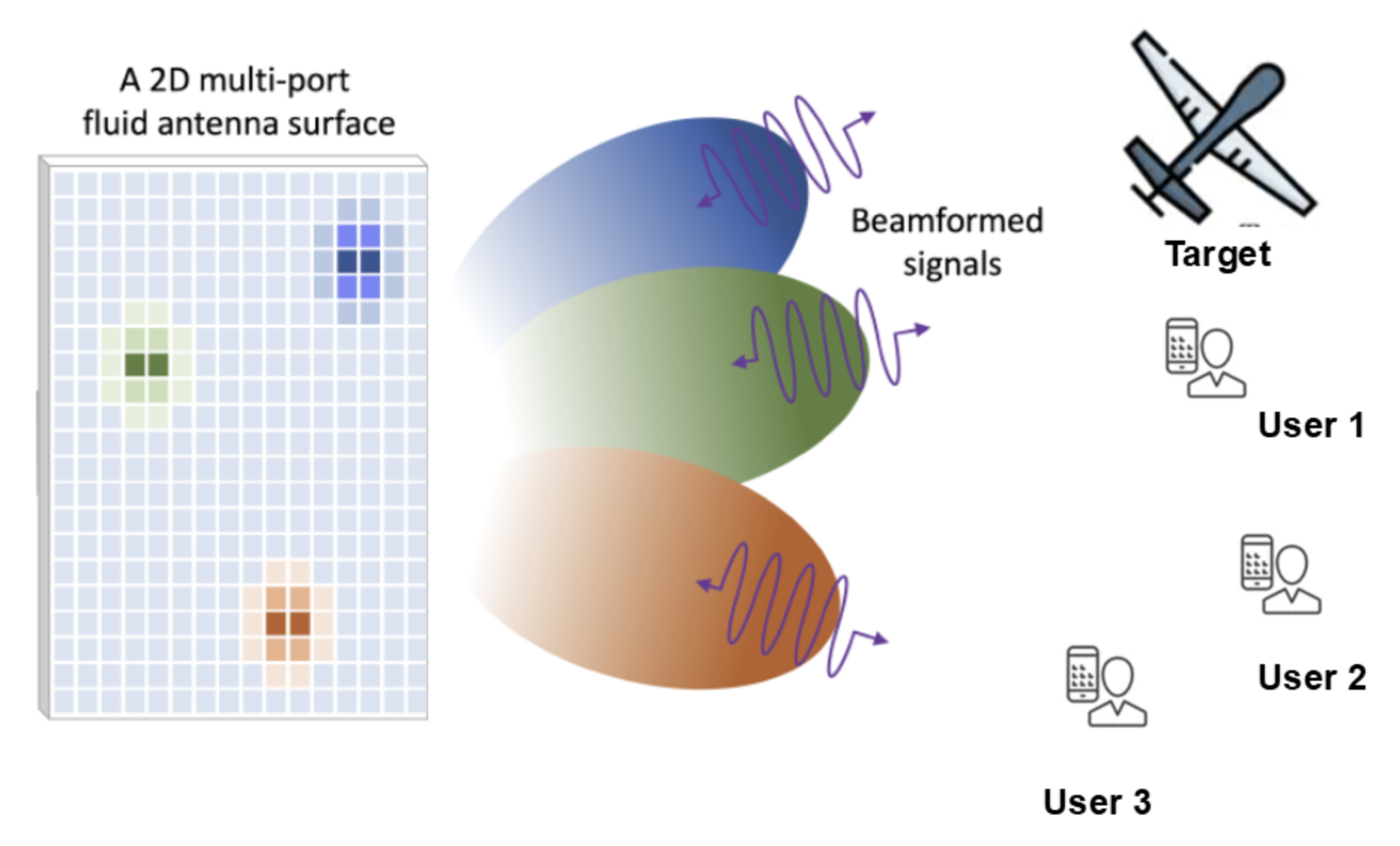}
\end{centering}
\caption{The FAS-assisted ISAC systems.}\label{fig:system}
\vspace{-2mm}
\end{figure}

Consider an ISAC system in which a base station (BS) is equipped with an $N_{s}$-port FAS to serve $K$ communication users in the downlink, and has $N_{r}$ fixed receive antennas to sense a single target, as shown in Fig.~\ref{fig:system}. The FAS occupies a two-dimensional (2D) surface with an area $A_{s}=\lambda A_{s}^{x}\times\lambda A_{s}^{y}$, where $\lambda$ is the wavelength of the carrier frequency $f_{\ensuremath{c}}$. The FAS has $N_{s}=N_{s}^{x}\times N_{s}^{y}$ ports that distribute evenly over the space, on the $xy$-plane, with $N_{s}^{i}$ ports along $\lambda A_{s}^{i}$, $i\in\left\{ x,y\right\} $. For notational convenience, we use vectorization of the index coordinates so that the $\left(n_{x,i}^{s},n_{y,i}^{s}\right)$-th port, for $0\le n_{x,i}^{s}\le N_{s}^{x}$ and $0\le n_{y,i}^{s}\le N_{s}^{y}$ in the 2D space has an index $\left(n_{y,i}^{s}+\left(n_{x,i}^{s}-1\right)N_{s}^{y}\right)$. We assume that the BS only activates $n_{s}$ out of $N_{s}$ ports to generate the ISAC waveform. All the users' messages are private and need to be kept confidential to each other. Considering internal eavesdroppers, the target and any user in the system can be an eavesdropper. 

\subsection{Communication Performance}
Assuming a rich-scattering environment, we adopt Jake's model to characterize the spatial correlation between any two ports in the FAS. Specifically, the $i$-th port and the $j$-th port, for $1\le i,j\le N_{s}$, have their channel correlation given by $J_{i,j}^{s}=j_{0}\left(\frac{2\pi\text{\textgreek{D}}d_{i,j}}{\lambda}\right)$, where $j_{0}\left(.\right)$ denotes the zeroth-order spherical Bessel function. Besides, the distance between the $i$-th port and the $j$-th port is $\Delta d_{i,j}\triangleq\sqrt{\left(\frac{\left|n_{x,i}^{s}-n_{x,j}^{s}\right|}{N_{s}^{x}-1}A_{s}^{x}\right)^{2}+\left(\frac{\left|n_{y,i}^{s}-n_{y,j}^{s}\right|}{N_{s}^{y}-1}A_{s}^{y}\right)^{2}}$ where $i=n_{y,i}^{s}+\left(n_{x,i}^{s}-1\right)N_{s}^{x}$ and $j=n_{y,j}^{s}+\left(n_{x,j}^{s}-1\right)N_{s}^{x}$. Obviously, the spatial correlation matrix $\mathbf{J}_{s}=\left[J_{i,j}^{s}\right]_{i,j\in\left[1,N_{s}\right]}$ is a symmetric matrix, and can be expressed also by, $\mathbf{J}_{s}=\mathbf{\Upsilon}\mathbf{\Lambda}\mathbf{\Upsilon}^{T}$, where $\mathbf{\Upsilon}\in\ensuremath{\mathbb{C}}^{N_{s}\times N_{s}}$ is a unitary matrix containing the eigenvectors of $\mathbf{J}_{s}$ and the diagonal matrix $\mathbf{\Lambda}$ consists of the eigenvalues of $\mathbf{J}_{s}$ in descending order. Following \cite{G10_new2024MIMO-FAS}, the channel from the BS to the $k$-th user can be written as
\begin{equation}
\mathbf{h}_{k}^{H}=\sqrt{l_{k}}\mathbf{g}_{k}^{H}\ensuremath{\bm{\Lambda}^{\frac{1}{2}}}\boldsymbol{\Upsilon}^{T},
\end{equation}
where $l_{k}=d_k^{-m}$ denotes the large-scale path loss, with $d_k$ being the distance between the BS and the user and $m$ being the path-loss exponent, and $\mathbf{g}_{k}\in\ensuremath{\mathbb{C}}^{N_{s}\text{\texttimes}1}$ is the small-scale fading channel vector whose elements follow the standard complex Gaussian distribution, i.e., $\ensuremath{{\cal CN}(0,1)}$. Therefore, we can define ${\bf H} \in \mathbb{C}^{K \times N_s}$ as the channel matrix between the BS and all $K$ users, given a set of selected ports $\mathcal{S} \subseteq \{1,2,\ldots,N_s\}$ with $|\mathcal{S}| = n_s$ so that $n_s$ specifies the number of selected ports. Also, the sub-channel matrix can be defined as  ${\bf H}_{\mathcal{S}}={\bf H}(:,\mathcal{S})$ which extracts the columns of ${\bf H}$ corresponding to the indices in $\mathcal{S}$. With $n_{s}$ distinct ports activated, the received signal at user $k$ can be written as 
\begin{equation}
y_{k}=\sqrt{P}\mathbf{h}_{k}^{H}\boldsymbol{\Pi}_{n_{s}}\mathbf{w}_{k}s_{k}+\sqrt{P} \sum_{i=1\atop i\neq k}^K \mathbf{h}_{k}\boldsymbol{\Pi}_{n_{s}}\mathbf{w}_{i}s_{i}+n_{k},
\end{equation}
where $\boldsymbol{\Pi}_{n_{s}}=\left[\mathbf{z}_{1},\dots,\mathbf{z}_{n_{s}}\right]$ is the port activation matrix with $\mathbf{z}_{i}$ being the $i$-th column of an $N_{s}\times N_{s}$ identity matrix, $s_{k}$ is user $k$'s signal with unit variance, $\mathbf{w}_{k}$ is the $n_{s}\times1$ precoding vector of user $k$, and $n_{k}$ denotes the additive white Gaussian noise (AWGN) at the user, i.e., $n_{k}\sim\mathcal{CN}\left(\text{0, }\sigma_{k}^{2}\right)$. Therefore, the received SINR at the user is given by 
\begin{equation}\label{eq:3}
\gamma_{k}=\frac{P\left|\mathbf{h}_{k}^{H}\boldsymbol{\boldsymbol{\Pi}}_{n_{s}}\mathbf{w}_{k}\right|^{2}}{P \sum_{i=1\atop i\neq k}^K \left|\mathbf{h}_{k}^{H}\boldsymbol{\boldsymbol{\Pi}}_{n_{s}}\mathbf{w}_{i}\right|^{2}+\sigma_{k}^{2}}.
\end{equation}
Therefore, the rate at user $k$ can be found as
\begin{equation}
R_{k}=\log_{2}\left(1+\gamma_k\right).
\end{equation}

\subsection{Radar Performance}
Radar systems operate by radiating an electromagnetic signal into a region and detecting the echo returned from the reflecting target. Following \cite{Ref1,Ref2}, the echo signal from the line-of-sight (LOS) path is exploited for the sensing. Thus, the received signal at the BS can be written as
\begin{equation}
\mathbf{y}_{b}=\sqrt{P}\mathbf{G}\boldsymbol{\boldsymbol{\Pi}}_{n_{s}}\mathbf{W}\mathbf{s}+\mathbf{z}_{c}+\mathbf{n}_{b},
\end{equation}
where $\mathbf{s}$ denotes the user message vector, $\mathbf{G}=\alpha\mathbf{a}_{r}\mathbf{a}_{t}^{H}$ is an $N_{r}\times N_{s}$ target response matrix at the BS, $\alpha$ is the amplitude of the target which contains the round-trip path-loss and the radar cross-section of the target, $\mathbf{a}_{t}=\left[e^{j\frac{2\pi}{\lambda}\triangle_{t}^{1}},\dots,e^{j\frac{2\pi}{\lambda}\triangle_{t}^{N_{s}}}\right]^{T}$ and $\mathbf{a}_{r}=\left[e^{j\frac{2\pi}{\lambda}\triangle_{r}^{1}},\dots,e^{j\frac{2\pi}{\lambda}\triangle_{r}^{N_{r}}}\right]^{T}$ are the associated transmit and receive array steering vectors, respectively, $\triangle_{m}^{n}$ is the propagation distance difference of the LOS link between the $n$-th port and the reference port,  $\mathbf{z}_{c}$ is Gaussian interference at the BS with the covariance matrix $\mathbf{R}_{c}$, and $\mathbf{n}_{b}$ is the disturbance in the receiver (accounting for the internal thermal noise, the sky noise, external disturbance, clutter, etc.) \cite{Spectral}. Given a set of selected ports $\mathcal{S} \subseteq \{1,2,\ldots,N_s\}$ with $|\mathcal{S}| = n_s$, the sub-channel matrix is defined as ${\bf G}_\mathcal{S} ={\bf G}(:,\mathcal{S})$, which extracts the columns of ${\bf G}$ corresponding to the indices in $\mathcal{S}$. 

The radar receiver uses a filter $\mathbf{w}_{r}$ to reduce the interference and noise. Then the filtered signal can be written by
\begin{equation}
\tilde{y}_{b}=\sqrt{P}\mathbf{w}_{r}^{H}\mathbf{G}\boldsymbol{\Pi}_{n_{s}}\mathbf{W}\mathbf{s}+\mathbf{w}_{r}^{H}\mathbf{z}_{c}+\mathbf{w}_{r}^{H}\mathbf{n}_{b}.
\end{equation}
Here, we propose to adopt the minimum variance distortion-less response (MVDR) beamformer, where $\mathbf{w}_{r}=\frac{\mathbf{\tilde{R}^{-1}}\,\mathbf{a}_{r}}{\mathbf{a}_{r}^{H}\,\mathbf{\tilde{R}}^{-1}\,\mathbf{a}_{r}}$, and $\tilde{\mathbf{R}}=\mathbf{R}_{c}+\mathbf{I}_{N_{r}}\sigma_{b}^{2}$ \cite{mvdr1}. As a result, the output SINR of the radar system with the fixed MVDR beam-former and uncorrelated interference is given by
\begin{equation}\label{eq:17-1}
\gamma_{b}=\frac{P\mathbf{w}_{r}^{H}\mathbf{G}\boldsymbol{\boldsymbol{\Pi}}_{n_{s}}\mathbf{W}\mathbf{s}\mathbf{s}^{H}\mathbf{W}^{H}\boldsymbol{\boldsymbol{\Pi}}_{n_{s}}^{H}\mathbf{G}^{H}\mathbf{w}_{r}}{\mathbf{w}_{r}^{H}\left(\mathbf{R}_{c}+\mathbf{I}_{N_{r}}\sigma_{b}^{2}\right)\mathbf{w}_{r}}.
\end{equation}

\subsection{Secrecy Performance}
In our scenario, the eavesdropper can be internal, i.e., the target or any other user in the system. Hence, each user's message is confidential and should be transmitted securely. In case the target acts as a potential eavesdropper. The eavesdropping
SINR on the $k$-th legitimate user is given by
\begin{equation}
\gamma_{e}^{r}=\frac{P\left|\mathbf{a}_{t}\boldsymbol{\boldsymbol{\Pi}}_{n_{s}}\mathbf{w}_{k}\right|^{2}} 
{ \sum_{i=1\atop i\neq k}^K P\left|\mathbf{a}_{t}\boldsymbol{\boldsymbol{\Pi}}_{n_{s}}\mathbf{w}_{i}\right|^{2}+\sigma_{r}^{2}}
\end{equation}
and the eavesdropper rate is 
\begin{equation}
R_{e}^{r}=\log_{2}\left(1+\gamma_{e}^{r}\right).
\end{equation}

For the case any other user $i$ in the system acts as a potential eavesdropper, the eavesdropping SINR of user $i$ to eavesdrop the $k$-th legitimate user is given by
\begin{equation}\label{eq:3-1}
\gamma_{e}^{i}=\frac{P\left|\mathbf{h}_{i}^{H}\boldsymbol{\boldsymbol{\Pi}}_{n_{s}}\mathbf{w}_{k}\right|^{2}}{P 
\sum_{j=1\atop j\neq k,i}^K\left|\mathbf{h}_{i}^{H}\boldsymbol{\boldsymbol{\Pi}}_{n_{s}}\mathbf{w}_{j}\right|^{2}+\sigma_{i}^{2}}
\end{equation}
and the rate is written as
\begin{equation}
R_{e}^{i}=\log_{2}\left(1+\gamma_{e}^{i}\right).
\end{equation}

Now, the secrecy rate of user $k$ can be defined as 
\begin{equation}
R_{s}^{k}=\left[R_{k}-\max\left(R_{e}^{r},R_{e}^{i}\right)\right]^{+}, \forall i,
\end{equation}
where $[a]^+=\max(0,a)$. Consequently, the sum secrecy rate can be found as 
\begin{equation}
R_{s}=\sum_{k=1}^K R_{s}^{k}.
\end{equation}

\section{Problem Formulation\label{sec:Problem-Formulation}}
The joint optimization problem of port selection and precoding to maximize the sum secrecy rate under the sensing and power constraints can be formulated as
\begin{eqnarray}
\underset{\boldsymbol{\boldsymbol{\Pi}}_{n_{s}}\mathbf{W}}{\max\,} & R_s\nonumber \\
\mbox{s.t.} &\gamma_{b}\geq\zeta & \left(C.1\right)\nonumber \\
 & \left[\boldsymbol{\boldsymbol{\Pi}}_{n_{s}}\right]_{l}\in\left\{ \mathbf{z}_{1},\dots,\mathbf{z}_{N_{s}}\right\}  & \left(C.2\right)\nonumber \\
 & \left[\boldsymbol{\boldsymbol{\Pi}}_{n_{s}}\right]_{l}\neq\left[\boldsymbol{\boldsymbol{\Pi}}_{n_{s}}\right]_{k} & \left(C.3\right)\nonumber \\
 & {\rm Tr}\left(\mathbf{W}\mathbf{W}^{H}\right)\leq P, & \left(C.4\right)\label{eq:main}
\end{eqnarray}
where $\zeta$ is the worst-case achieved radar sensing SINR. The first constraint ($C.1$) is set to achieve the radar sensing SINR, while the second and third constraints aim to enforce the $n_{s}$ activated ports to be distinct. The last constraint is the power constraint. Obviously, (\ref{eq:main}) is a non-convex optimization problem due to the non-convexity of the objective function. In addition, the precoder design and port selection are mutually coupled, which makes the joint optimization more challenging. Although one can employ exhaustive search to find the optimal port combination, which results in ${N_s\choose n_s}$ non-convex problems, its complexity is prohibitively high. As an alternative, existing works propose some approximate algorithms, which usually separate the port selection and the precoder optimization into two optimization problems and then approximate the port selection as some tractable problems. 

\section{Proposed Solutions \label{sec:Proposed-Solutions}}
In this section, we present two efficient methods to solve the problem. First, the problem is simplified into a more tractable yet equivalent form by invoking the FP framework. Then the beamforming matrix \textbf{$\mathbf{W}$} and the best ports are updated in an alternating manner. In the second method, ZF beamforming is implemented, and then greedy and trace inverse minimization schemes are proposed to obtain sub-optimal solutions.

\subsection{Joint Precoding and Port Selection (JPPS) via FP\label{sub:Joint}}
Here, we present a rigorous formulation and solution to the secrecy rate maximization problem in MIMO-ISAC systems using the FP framework considering multiple eavesdroppers including the sensing target and legitimate users. 

To begin, define the worst-case eavesdropper SINR of user $k$ as 
\begin{equation}
\theta_{k}=\max\left(\gamma_{e}^{r},\max_{i\ne k}\gamma_{e}^{i}\right).
\end{equation}
Thus, the secrecy rate for user $k$ can be expressed as 
\begin{equation}
R_{s}^{k}=\log_{2}\left(\frac{1+\gamma_{k}}{1+\theta_{k}}\right).
\end{equation}
Then the sum secrecy rate maximization problem becomes
\begin{eqnarray}
\underset{\boldsymbol{\boldsymbol{\Pi}}_{n_{s}},\mathbf{W}}{\max} & \sum_{k=1}^{K}\log_{2}\left(\frac{1+\gamma_{k}}{1+\theta_{k}}\right)\nonumber \\
\mbox{s.t.} & \gamma_{b}(\mathbf{W})\geq\zeta\nonumber \\
 & \left[\boldsymbol{\boldsymbol{\Pi}}_{n_{s}}\right]_{l}\in\left\{ \mathbf{z}_{1},\dots,\mathbf{z}_{N_{s}}\right\} \nonumber \\
 & \left[\boldsymbol{\boldsymbol{\Pi}}_{n_{s}}\right]_{l}\neq\left[\boldsymbol{\boldsymbol{\Pi}}_{n_{s}}\right]_{k}\nonumber \\
 & {\rm Tr}(\mathbf{W}\mathbf{W}^{H})\leq P_{\max}.\label{eq:16-1}
\end{eqnarray}

By applying the FP method, the rate at user $k$ and the worst-case eavesdropper of user $k$ can be expressed, respectively, as functions of the auxiliary variables $u_{k}$ and $v_{k}$ so that
\begin{multline}
\log_{2}(1+\gamma_{k})=\log_{2}(1+u_{k})-u_{k}\\
+\left(1+u_{k}\right)\left[2{\rm Re}\left(\delta_{k}a_{k}\right)-\left|\delta_{k}\right|^{2}b_{k}\right]
\end{multline}
and
\begin{multline}\label{eq:uvdb}
\log_{2}(1+\theta_{k})=\log_{2}(1+v_{k})-v_{k}\\
+\left(1+v_{k}\right)\left[2{\rm Re}\left(\beta_{k}c_{k}\right)-\left|\beta_{k}\right|^{2}e_{k}\right],
\end{multline}
where $u_{k}$, $v_{k}$, $\delta_{k}$ and $\beta_{k}$ are the auxiliary variables, $a_{k}=\mathbf{h}_{k}^{H}\boldsymbol{\boldsymbol{\Pi}}_{n_{s}}\mathbf{w}_{k}$, $b_{k}=\sum_{i=1}^K\left|\mathbf{h}_{k}^{H}\boldsymbol{\boldsymbol{\Pi}}_{n_{s}}\mathbf{w}_{i}\right|^{2}+\sigma_{k}^{2}$, $c_{k}=\mathbf{g}_{k}^{H}\boldsymbol{\boldsymbol{\Pi}}_{n_{s}}\mathbf{w}_{k}$, $e_{k}=\sum_{i=1}^K\left|\mathbf{g}_{k}^{H}\boldsymbol{\boldsymbol{\Pi}}_{n_{s}}\mathbf{w}_{k}\right|^{2}+\sigma_{e}^{2}$, and $\mathbf{g}_{k}$ represents the worst-case eavesdropper for user $k$. Now the problem in (\ref{eq:16-1}) can be equivalently transformed to
\begin{eqnarray}
\max_{\boldsymbol{\boldsymbol{\Pi}}_{n_{s}},\mathbf{W},\mathbf{u},\atop \mathbf{v},\boldsymbol{\delta},\boldsymbol{\beta}}
& \hspace{-4mm}
\begin{array}{l}
\sum_{k=1}^{K}\left(\log_{2}(\frac{1+u_{k}}{1+v_{k}})-u_{k}\right.\\
+\left(1+u_{k}\right)\left[2{\rm Re}\left(\delta_{k}a_{k}\right)-\left|\delta_{k}\right|^{2}b_{k}\right]\\
\left.+v_{k}-\left(1+v_{k}\right)\left[2{\rm Re}\left(\beta_{k}c_{k}\right)-\left|\beta_{k}\right|^{2}e_{k}\right]\right)
\end{array}\nonumber\\
\mbox{s.t.} & \hspace{-4mm}\gamma_{b}(\mathbf{W})\geq\zeta\nonumber \\
 & \hspace{-4mm}\left[\boldsymbol{\boldsymbol{\Pi}}_{n_{s}}\right]_{l}\in\left\{ \mathbf{z}_{1},\dots,\mathbf{z}_{N_{s}}\right\} \nonumber \\
 & \hspace{-4mm}\left[\boldsymbol{\boldsymbol{\Pi}}_{n_{s}}\right]_{l}\neq\left[\boldsymbol{\boldsymbol{\Pi}}_{n_{s}}\right]_{k}\nonumber \\
 & \hspace{-4mm}{\rm Tr}(\mathbf{W}\mathbf{W}^{H})\leq P_{\max}.\label{eq:19}
\end{eqnarray}

The optimal values of $u_{k}$, $v_{k}$, $\delta_{k}$ and $\beta_{k}$ can be given by $u_{k}=\frac{\left|a_{k}\right|^{2}}{\sum_{i=1\atop i\neq k}^K{\sum}|\mathbf{h}_{k}^{H}\mathbf{w}_{j}|^{2}+\sigma_{k}^{2}}$, $v_{k}=\frac{\left|c_{k}\right|^{2}}{\sum_{i=1\atop i\neq k}^K|\mathbf{g}_{k}^{H}\mathbf{w}_{j}|^{2}+\sigma_{e}^{2}}$,
$\delta_{k}=a_{k}b_{k}^{-1}$ and $\beta_{k}=c_{k}e_{k}^{-1}$. To decouple the variables in (\ref{eq:19}), we propose to optimize each variable iteratively with other variables fixed. Since the conditionally optimal $u_{k}$, $v_{k}$, $\delta_{k}$ and $\beta_{k}$ are already presented, we intend to develop the iterative design of $\boldsymbol{\boldsymbol{\Pi}}_{n_{s}}$ and $\mathbf{W}$, which is discussed next.

\subsubsection{Optimizing the Beamforming Matirx $\mathbf{W}$}
The marginal problem for $\mathbf{W}$ can be expressed as
\begin{eqnarray}
\underset{\mathbf{W}}{\max} & 2{\rm Re}\left\{ \left({\rm Tr}\left(\mathbf{W}^{H}\mathbf{C}_{1}\right)\right)\right\} -{\rm Tr}\left(\mathbf{W}^{H}\mathbf{D}_{1}\mathbf{W}\right)\nonumber \\
 & -2{\rm Re}\left\{ \left({\rm Tr}\left(\mathbf{W}^{H}\mathbf{C}_{2}\right)\right)\right\} + {\rm Tr}\left(\mathbf{W}^{H}\mathbf{D}_{2}\mathbf{W}\right)\nonumber \\
\textrm{s.t.} & {\rm Tr}\left(\mathbf{W}^{H}\mathbf{D}_{3}\mathbf{W}\right)\geq\varrho\nonumber \\
 & {\rm Tr}(\mathbf{W}\mathbf{W}^{H})\leq P_{\max},\label{eq:solveW}
\end{eqnarray}
where 
\begin{equation}
\left\{\begin{aligned}
\mathbf{C}_{1}&=\left[\left(1+u_{1}\right)\delta_{1}\mathbf{h}_{1},\dots,\left(1+u_{K}\right)\delta_{K}\mathbf{h}_{K}\right],\\
\mathbf{D}_{1}&=\sum_{k=1}^{K}\left(1+u_{k}\right)\left|\delta_{k}\right|^{2}\mathbf{h}_{k}\mathbf{h}_{k}^{H},\\
\mathbf{C}_{2}&=\left[\left(1+v_{1}\right)\beta_{1}\mathbf{g}_{1},\dots,\left(1+v_{K}\right)\beta_{K}\mathbf{g}_{K}\right],\\
\mathbf{D}_{2}&=\sum_{k=1}^{K}\left(1+v_{k}\right)\left|\beta_{k}\right|^{2}\mathbf{g}_{k}\mathbf{g}_{k}^{H},\\
\mathbf{D}_{3}&=\boldsymbol{\boldsymbol{\Pi}}_{n_{s}}\mathbf{a}_{t}\mathbf{a}_{t}^{H}\boldsymbol{\boldsymbol{\Pi}}_{n_{s}}^{H},\\
\varrho&=\zeta\frac{\mathbf{w}_{r}^{H}\left(\mathbf{R}_{c}+\mathbf{I}_{N_{r}}\sigma_{b}^{2}\right)\mathbf{w}_{r}}{P\alpha^{2}}.
\end{aligned}\right.
\end{equation}

To solve (\ref{eq:solveW}), we use SCA or the majorization-minimization (MM) method, which approximates the non-convex parts locally by convex surrogates. At iteration $t$, we approximate the problematic convex parts by its first-order Taylor expansion around current $\mathbf{W}^{(t)}$. Thus, each sub-problem becomes convex. First-order Taylor expansion of ${\rm Tr}\left(\mathbf{W}^{H}\mathbf{D}_{i}\mathbf{W}\right)$ around $\mathbf{W}^{(t)}$ after removing the constants can be written as
\begin{equation}\label{eq:linear-con}
{\rm Tr}\left(\mathbf{W}^{H}\mathbf{D}_{i}\mathbf{W}\right)\approx2{\rm Re}\left\{ {\rm Tr}\left(\left(\mathbf{D}_{i}\mathbf{W}^{\left(t\right)}\right)^{H}\left(\mathbf{W}-\mathbf{W}^{\left(t\right)}\right)\right)\right\}.
\end{equation}
Accordingly, at each iteration $t$, we need to solve
\begin{eqnarray}
\underset{\mathbf{W}}{\max} & 2{\rm Re}\left\{ {\rm Tr}\left(\mathbf{W}^{H}\left(\mathbf{C}_{1}-\mathbf{C}_{2}+\mathbf{D}_{2}\mathbf{W}^{\left(t\right)}-\mathbf{D}_{1}\mathbf{W}^{\left(t\right)}\right)\right)\right\} \nonumber \\
\mbox{s.t.} & \left\Vert \mathbf{W}\right\Vert _{F}\leq\sqrt{P_{\max}}\nonumber \\
 & 2{\rm Re}\left\{ {\rm Tr}\left(\mathbf{W}^{H}\mathbf{D}_{3}\mathbf{W}^{\left(t\right)}\right)\right\} \geq\varrho.\label{eq:24-1}
\end{eqnarray}

Now the problem can be solved using CVX at each iteration until convergence. The steps are stated in Algorithm \ref{alg:sca}.

\begin{algorithm}
\caption{SCA for Beamforming Optimization}\label{alg:sca}
\hspace*{0.02in} \textbf{Input:} Initialize $t=0$ and choose $\mathbf{W}^{\left(0\right)}$ randomly s.t.~($C.4$).\\
\hspace*{0.02in} \textbf{Output:} Optimized $\mathbf{W}$.
\begin{algorithmic}[1]
\State Step 1: Compute linear approximation of the objective terms at $\mathbf{W}^{\left(t\right)}$, see the objective function in (\ref{eq:24-1}).
\State Step 2: Compute linear approximation of radar constraint at $\mathbf{W}^{\left(t\right)}$ using (\ref{eq:linear-con}).
\State Step 3: Solve the convex sub-problem (\ref{eq:24-1}).
\State Step 4: Update: $\mathbf{W}^{\left(t+1\right)}=\mathbf{W}$ and $t\leftarrow t+1$.
\State Step 5: Check convergence and if not, go back to Step 1.
\State \Return $\mathbf{W}$.
\end{algorithmic}
\end{algorithm}

\begin{remark}
Without the radar constraints, the problem will become a standard convex quadratic optimization problem whose solution is $\mathbf{W}=\left(\mathbf{D}_{1}+\lambda\mathbf{I}\right)^{-1}\mathbf{D}$ where the regularizer $\lambda$ is chosen such that the complementarity slackness condition, i.e., $\left(\lambda{\rm Tr}\left(\mathbf{W}\mathbf{W}^{H}\right)-P\right)=0$, is satisfied. 
\end{remark}

\subsubsection{Optimizing the Ports Positions $\boldsymbol{\boldsymbol{\Pi}}_{n_{s}}$}
The marginal problem for the ports selection can be expressed as
\begin{eqnarray}
\underset{\boldsymbol{\Pi}_{n_{s}}}{\max} & 
\begin{array}{l}
\sum_{k=1}^{K}\left[\left(1+u_{k}\right)\left[2{\rm Re}\left(\delta_{k}a_{k}\right)-\left|\delta_{k}\right|^{2}b_{k}\right]\right]\\
-\left(1+v_{k}\right)\left[2{\rm Re}\left(\beta_{k}c_{k}\right)-\left|\beta_{k}\right|^{2}e_{k}\right]
\end{array}\nonumber\\
\mbox{s.t.} & \left[\Pi_{n_{s}}\right]_{l}\in\left\{ \mathbf{z}_{1},\dots,\mathbf{z}_{N_{s}}\right\} \nonumber \\
 & \left[\boldsymbol{\boldsymbol{\Pi}}_{n_{s}}\right]_{l}\neq\left[\boldsymbol{\boldsymbol{\Pi}}_{n_{s}}\right]_{k}.\label{eq:24}
\end{eqnarray}

This is a combinatorial optimization problem over the port activation matrix $\boldsymbol{\boldsymbol{\Pi}}_{n_{s}}$, selecting the best subset of $n_{s}$ \LyXZeroWidthSpace{} ports out of $N_{s}$ available to maximize a secrecy-aware utility function derived from FP. The problem is non-convex and discrete, as the optimization domain is the set $\mathcal{P}_{N,n_{s}}\text{\ensuremath{\subset}}\left\{ 0,1\right\} ^{N_{s}}$. Let us define $\mathbf{r}\in\left\{ 0,1\right\} ^{N_{s}}$ as a binary selection vector where $r_{n}=1$ if port $n$ is selected. Thus the constraint $\sum_{n=1}^{N_s}r_{n}=n_{s}$ ensures that there are exactly $n_{s}$ active ports. Now problem (\ref{eq:24}) can be reformulated as
\begin{equation}
\max_{\mathbf{r}\text{\ensuremath{\in}}\left\{ 0,1\right\} ^{N_{s}}\atop \left\Vert \mathbf{r}\right\Vert _{0}=n_{s}} \mathcal{U}\left(\mathbf{r};\mathbf{W},\mathbf{u},\boldsymbol{\delta},\mathbf{v},\boldsymbol{\beta}\right),
\end{equation}
in which $\mathcal{U}(\cdot)$ is the total FP-based secrecy utility function. By using the facts that $\mathbf{h}_{k}^{H}\mathbf{w}_{k}=\sum_{n=1}^{N_{s}}h_{k}^{H}\left(n\right)w_{k,n}$, ${\rm Re}\left(\delta_{k}\mathbf{h}_{k}^{H}\mathbf{w}_{k}\right)=\sum_{n=1}^{N}{\rm Re}\left(\delta_{k}h_{k}^{H}\left(n\right)w_{k,n}\right)$, $\left|\mathbf{h}_{k}^{H}\mathbf{w}_{k}\right|^{2}=\left|\sum_{n=1}^{N_{s}}h_{k}^{H}\left(n\right)w_{k,n}\right|^{2}=\underset{n}{\sum}\underset{n'}{\sum}w_{j,n}w_{j,n'}^{H}h_{k}^{H}\left(n\right)h_{k}\left(n'\right)=\underset{n,n'}{\sum}W_{n,n'}h_{k}^{H}\left(n\right)h_{k}\left(n'\right)$, we can write
\begin{equation}\label{eq:26}
\mathcal{U}\left(\mathbf{r};\mathbf{W},\mathbf{u},\boldsymbol{\delta},\mathbf{v},\boldsymbol{\beta}\right)=\sum_{k=1}^{K}\sum_{n=1}^{N}r_{n}\left(\Phi_{k,n}^{h}\left(\mathbf{r}\right)-\Phi_{k,n}^{g}\left(\mathbf{r}\right)\right),
\end{equation}
where 
\begin{multline}
\Phi_{k,n}^{h}\left(\mathbf{r}\right)=\sum_{k=1}^{K}\sum_{n=1}^{N_{s}}\left[2{\rm Re}\left(\left(1+u_{k}\right)\delta_{k}h_{k}^{H}\left(n\right)w_{k,n}\right)\right.\\
-\left|\delta_{k}\right|^{2}\left(1+u_{k}\right)h_{k}^{H}\left(n\right)\underset{n'\neq n}{\sum}r_{n'}W_{n,n'}h_{k}\left(n'\right)\\
\left.-\left|\delta_{k}\right|^{2}\left(1+u_{k}\right)W_{n,n}\left|h_{k}\left(n\right)\right|^{2}\right]
\end{multline}
and 
\begin{multline}
\Phi_{k,n}^{g}\left(\mathbf{r}\right)=\sum_{k=1}^{K}\sum_{n=1}^{N_{s}}\left[2{\rm Re}\left(\left(1+v_{k}\right)\beta_{k}g_{k}^{H}\left(n\right)w_{k,n}\right)\right.\\
-\left|\beta_{k}\right|^{2}\left(1+v_{k}\right)g_{k}^{H}\left(n\right)\underset{n'\neq n}{\sum}r_{n'}W_{n,n'}g_{k}\left(n'\right)\\
\left.-\left|\beta_{k}\right|^{2}\left(1+v_{k}\right)W_{n,n}\left|g_{k}\left(n\right)\right|^{2}\right].
\end{multline}

To avoid discrete search, we relax $\mathbf{r}\in\left\{ 0,1\right\} ^{N_{s}}$ to $\mathbf{r}\in[0,1]^{N_{s}}$, and reformulate the problem as
\begin{equation}
\underset{\mathbf{r}\text{\ensuremath{\in}}[0,1]^{N_{s}}}{\max}\mathcal{U}\left(\mathbf{r}\right)~
\mbox{s.t.}~\sum_{n=1}^{N_{s}}r_{n}=n_{s}.
\end{equation}
This convex relaxation of the port selection problem can be solved using CVX and integrated into the FP-based precoding loop. After solving this convex surrogate, we choose the $n_{s}$ largest $r_{n}$ values to construct the final binary vector. 

To reduce the complexity, we can consider a greedy heuristic port selection method. Let us define
\begin{equation}
\text{\textgreek{G}}_{n}=\sum_{k=1}^{K}\left|h_{k}\left(n\right)\right|\left(1+u_{k}\right)\left|\delta_{k}\right|-\left(1+v_{k}\right)\left|\beta_{k}\right|\left|g_{k}\left(n\right)\right|,
\end{equation}
which acts as a utility score per port derived from the FP structure. We then solve the problem
\begin{equation}
\max_{\mathbf{r}\text{\ensuremath{\in}}\left\{ 0,1\right\} ^{N_{s}}\atop \left\Vert \mathbf{r}\right\Vert _{0}=n_{s}}\sum_{n=1}^{N_s}\text{\textgreek{G}}_{n},
\end{equation}
which is a simple knapsack problem, and the optimal solution is obtained by selecting the top $n_{s}$ ports with the largest $\text{\textgreek{G}}_{n}$. 

Finally, all steps to obtain the optimal precoder and active ports via the FP-aware utility are described in Algorithm \ref{alg:jpps}. 

\begin{algorithm}
\caption{JPPS via the FP Scheme}\label{alg:jpps}
\hspace*{0.02in} \textbf{Input:} Initialize $t=0$, select initial active ports $\mathcal{S}^{\left(0\right)}\subset\left\{ 1,2,\dots,N_{s}\right\}$, with $\left|\mathcal{S}^{\left(0\right)}\right|=n_{s}$ and construct initial beamformer $\mathbf{W}^{(0)}\in C^{n_{s}\LyXZeroWidthSpace\times K}$.\\
\hspace*{0.02in} \textbf{Output:} Optimized $\mathcal{S}$.
\begin{algorithmic}[1]
\State Step 1: Extract the sub-channels for active ports $\mathbf{H}_{\mathcal{S}}=\mathbf{H}\left(:,\mathcal{S}^{\left(t\right)}\right)$, $\mathbf{G}_{\mathcal{S}}=\mathbf{G}\left(:,\mathcal{S}^{\left(t\right)}\right)$.
\State Step 2: Compute the auxiliary variables for each user $u_{k},\delta_{k},v_{k},\beta_{k}$ according to the definitions after (\ref{eq:uvdb}).
\State Step 3: Compute the port utility function $\mathcal{U}$ in (\ref{eq:26}).
\State Step 4: Choose $n_{s}$ ports with the highest $\mathcal{U}$. 
\State Step 5: Set $\mathcal{S}^{\left(t+1\right)}$ accordingly.
\State Step 6: Update the sub-channels $\mathbf{H}_{\mathcal{S}}=\mathbf{H}\left(:,\mathcal{S}^{\left(t+1\right)}\right)$, $\mathbf{G}_{\mathcal{S}}=\mathbf{G}\left(:,\mathcal{S}^{\left(t+1\right)}\right)$.
\State Step 7: Solve the precoding problem (\ref{eq:24-1}) using CVX.
\State Step 8: Set $\mathcal{S}^{\left(t+1\right)}=\mathcal{S}^{\left(t\right)}$.
\State Step 8: If $\left\Vert \mathbf{W}^{\left(t+1\right)}-\mathbf{W}^{\left(t\right)}\right\Vert _{F}^{2}<\epsilon$ is not true, then set $t\leftarrow t+1$ and go back to Step 2.
\State \Return $\mathcal{S}$.
\end{algorithmic}
\end{algorithm}

\subsubsection{Complexity Analysis}
Each user requires $\mathcal{O}\left(n_{s}K\right)$ operations to compute all inner products, and thus the total cost is $\mathcal{O}\left(n_{s}K^{2}\right)$. We compute the utility function for each $n=1,\ldots,N_{s}$, with the cost per port bwing $\mathcal{O}\left(K\right)$ and for all ports being $\mathcal{O}\left(N_{s}K\right)$. Then select top $n_{s}$ $\mathcal{O}\left(N_{s}\log N_{s}\right)$. The total cost is $\mathcal{O}\left(KN_{s}+N_{s}\log N_{s}\right)$. The beamforming has complexity $\mathcal{O}\left(\left(n_{s}K\right)^{3}\right)$. Hence, the total per-iteration complexity is $\mathcal{O}\left(K^{2}n_{s}+KN_{s}+N_{s}\log N_{s}+\left(n_{s}K\right)^{3}\right)$. Note that $KN_{s}$ dominates when the port set is large, and $(n_{s}K)^{3}$ dominates when beamforming becomes bottleneck in large systems with $N_{s}\log N_{s}$ negligible compared to other terms. 

\subsection{A Simple Solution}
For simplicity and to provide full security across the users, here we propose to adopt the ZF precoder to design $\mathbf{W}$ and then the port selection can be designed to achieve secure transmission while achieving the required sensing SINR. Thus, the precoding matrix can be written as $\mathbf{W}=\,\frac{\mathbf{H}^{H}\left(\mathbf{\mathbf{H}}\mathbf{H}^{H}\right)^{-1}}{\sqrt{{\rm Tr}\left(\left(\mathbf{\mathbf{H}}\mathbf{H}^{H}\right)^{-1}\right)}}$. As such, the SINR expression in (\ref{eq:3}) can be rewritten as 
\begin{equation}\label{eq:8-1}
\gamma_{k}=\frac{P}{\sigma_{k}^{2}{\rm Tr}\left(\left(\mathbf{\mathbf{H}}\mathbf{H}^{H}\right)^{-1}\right)}
\end{equation}
and the sum-rate can be rewritten as
\begin{equation}
R_{\textrm{com}}=\sum_{i=1}^K\log_{2}\left(1+\gamma_{k}\right).
\end{equation}
In addition, the output SINR of the radar system is
\begin{multline}
\gamma_{b}=\\
\frac{\left|\alpha\right|^{2}P\mathbf{a}_{t}^{T}\boldsymbol{\boldsymbol{\Pi}}_{n_{s}}\mathbf{H}^{H}\left(\mathbf{\mathbf{H}}\mathbf{H}^{H}\right)^{-1}\mathbf{s}\mathbf{s}^{H}\mathbf{H}^{H}\left(\mathbf{\mathbf{H}}\mathbf{H}^{H}\right)^{-1}\boldsymbol{\boldsymbol{\Pi}}_{n_{s}}\mathbf{a}_{t}}{{\rm Tr}\left(\left(\mathbf{\mathbf{H}}\mathbf{H}^{H}\right)^{-1}\right)\left(\mathbf{w}_{r}^{H}\left(\mathbf{R}_{c}+\mathbf{I}_{N_{r}}\sigma_{b}^{2}\right)\mathbf{w}_{r}\right)}.
\end{multline}
On the other hand, the eavesdropping SINR on the $k$-th legitimate user can be written as 
\begin{equation}
\gamma_{e}^{r}=\frac{\left|\mathbf{a}_{t}\left[\mathbf{H}^{H}\left(\mathbf{\mathbf{H}}\mathbf{H}^{H}\right)^{-1}\right]_{k}\right|^{2}}{\sum_{i=1\atop i\neq k}^K\left|\mathbf{a}_{t}\left[\mathbf{H}^{H}\left(\mathbf{\mathbf{H}}\mathbf{H}^{H}\right)^{-1}\right]_{i}\right|^{2}}
\end{equation}
and the eavesdropper rate is given by
\begin{equation}
R_{e}=\log_{2}\left(1+\gamma_{e}^{r}\right).
\end{equation}
Now, the problem to find the optimal ports can be recast as
\begin{eqnarray}
\max_{\mathbf{\Pi_{ns}}} & \sum_{k=1}^{K}R_{k}^{s}\nonumber \\
\mbox{s.t.} & \gamma_{b}\geq\zeta,\nonumber \\
 & \boldsymbol{\boldsymbol{\Pi}}_{n_{s}}\in\mathcal{P}_{N_{s},n_{s}},\label{eq:solvePI}
\end{eqnarray}
where $\mathcal{P}_{N_{s},n_{s}}$is the set of all possible $n_{s}$-port activation matrices formed from $N_{s}$ candidates. To solve (\ref{eq:solvePI}), we introduce the greedy selection (GS) and trace inverse minimization methods which will be explained next.

\subsubsection{GS}
Unfortunately, finding the optimal ports requires an exhaustive search, which becomes impractical in a massive MIMO setting. Since exhaustive search is highly computationally intensive, we propose a greedy port selection algorithm to reduce the computational complexity. The proposed algorithm starts with the full set of ports and then removes one port per step from this set. In each step, select one port (column) that results in the smallest decrease in the objective function while ensuring that the constraint is satisfied. If the radar constraint is satisfied, the candidate port removal is acceptable (i.e., remove the worst port and update the set); if the radar constraint is not satisfied, i.e., removing the port would violate the radar SINR constraint, then the algorithm must skip this removal and keep the port in the active set. The algorithm terminates when $n_{s}$ columns/ports are selected. To reduce overhead, we can skip port removals that immediately violate the radar SINR. All steps of the proposed GS Algorithm are listed in Algorithm \ref{alg:gs}.

\begin{algorithm}
\caption{GS for Port Selection}\label{alg:gs}
\hspace*{0.02in} \textbf{Input:} Set of the selected ports: $\mathcal{S}\leftarrow\left\{ 1,2,\dots,N_{s}\right\}$ and set of the remaining ports: $\mathcal{S}_{active}=\mathcal{S}$.\\
\hspace*{0.02in} \textbf{Output:} Optimized $\boldsymbol{\Pi}_{n_s}$.
\begin{algorithmic}[1]
\State Step 1: For each $j\in\mathcal{S}_{active}$, temporarily remove $j$ so that $\mathcal{S}_{j}=\mathcal{S}_{active}\setminus{j}$.
\State Step 2: Form $\boldsymbol{\Pi}_{j}$ from $\mathcal{S}_{j}$.
\State Step 3: Compute secrecy rate: $R_{\rm sum}^{j}=\sum_{k=1}^{K}R_{k}^{s}\left(\boldsymbol{\boldsymbol{\Pi}}_{j}\right)$.
\State Step 4: Check if $\gamma_b \geq \zeta$. If the condition is satisfied, then proceed to Step 5. Otherwise, skip the removal of port $j$ and keep it in the active set and go to Step 7.
\State Step 5: Remove port $j^{*}={\rm argmin}_{j} R_{\rm sum}^{j}$.
\State Step 6: Update: $\mathcal{S}_{active}\leftarrow\mathcal{S}_{active}\text{\ensuremath{\setminus}}{j^{*}}$.
\State Step 7: If $\left|\mathcal{S}_{active}\right|=n_{s}$ is not true, go back to Step 1.
\State \Return $\boldsymbol{\Pi}_{n_s}$ based on $\mathcal{S}_{active}$.
\end{algorithmic}
\end{algorithm}

\emph{Complexity Analysis:} Computing the full secrecy rate requires computing the inverse of $\mathbf{H}^{H}\mathbf{\mathbf{H}}$, which has a complexity of $O(N_{s}^{3})$. For each port removal, the evaluation of the secrecy rate involves recalculating $\left(\mathbf{H}^{H}\mathbf{\mathbf{H}}\right)^{-1}$, which takes $O((N_{s}-i)^{3})$ for each iteration $i$. Thus, the total complexity is $O(N_{s}^{3})+\stackrel[i=0]{N_{s}-n_{s}}{\sum}O((N_{s}\text{\textminus}i)^{3})$. Using summation approximation, the total complexity simplifies to $O(N_{s}^{4}\text{\LyXZeroWidthSpace})$.


\subsubsection{Trace Inverse Minimization}
To simplify the algorithm further, it is observed that the term ${\rm Tr}\left(\left(\mathbf{\mathbf{H}}\mathbf{H}^{H}\right)^{-1}\right)$ has the greatest impact on the objective function. Thus, minimizing this term can help to increase the secrecy rate and the sensing SINR. Thus, to simplify the algorithm, we can consider
\begin{eqnarray}
\underset{\boldsymbol{\boldsymbol{\Pi}}_{n_{s}}}{\min} & {\rm Tr}\left(\left(\mathbf{\mathbf{H}}\mathbf{H}^{H}\right)^{-1}\right)\nonumber \\
\mbox{s.t.} & \gamma_{b}\geq\zeta,\nonumber \\
 & \boldsymbol{\boldsymbol{\Pi}}_{n_{s}}\in\mathcal{P}_{N_{s},n_{s}}.
\end{eqnarray}
This is now a minimum trace inverse problem, widely used in linear precoding and antenna selection problems. We adopt GS for trace inverse minmization (GS-TIM) algorithm, which iteratively selects the port that provides the largest marginal reduction in the objective value. The GS-TIM algorithm provides near-optimal performance and aligns with theoretical guarantees from sub-modular set function optimization. The steps of the GS-TIM are presented in Algorithm \ref{alg:gstim}.

\begin{algorithm}
\caption{GS-TIM}\label{alg:gstim}
\hspace*{0.02in} \textbf{Input:} Initialize: $\mathcal{S}_{0}\leftarrow\emptyset$.\\
\hspace*{0.02in} \textbf{Output:} Optimized $\boldsymbol{\Pi}_{n_s}$.
\begin{algorithmic}[1]
\For{$t=1$ to $n_s$}
\State For each $j\notin\mathcal{S}_{t-1}$, 
\State Step 1: Form the candidate set $\mathcal{S}_{t-1}\cup\{j\}$.
\State Step 2: Compute the cost 
$$\mathcal{C}_{j}={\rm Tr}\left(\left(\mathbf{H}_{\mathcal{S}_{t-1}\cup\{j\}}\mathbf{H}_{\mathcal{S}_{t-1}\cup\{j\}}^{H}\right)^{-1}\right).$$
\State Step 3: Select $j^{*}=\arg\min_{j}\mathcal{C}_{j}$ and $\mathcal{S}_{t}\leftarrow\mathcal{S}_{t-1}\cup\{j^{*}\}$.
\EndFor
\State \Return $\boldsymbol{\Pi}_{n_s}=\mathcal{S}_{n_{s}}$.
\end{algorithmic}
\end{algorithm}

\emph{Complexity Analysis:} Each iteration requires a matrix inverse of size $K\times K$, leading to an overall complexity of $\mathcal{O}(n_{s}(N_{s}-n_{s})K^{3})$ which is tractable for moderate systems.

However, still, calculating the matrix inverse has very high complexity. In cases if $\mathbf{H}$ has full-rank, we can leverage the properties of the trace and matrix inverses, particularly using SVD of $\mathbf{\mathbf{H}}$ and use the SVD-TIM algorithm instead. The SVD of $\mathbf{\mathbf{H}}$ is given by $\mathbf{\mathbf{H}}=\mathbf{U}\boldsymbol{\varSigma}\mathbf{V}^{H}$, where $\mathbf{U}$ and $\mathbf{V}$ are unitary matrices, and $\boldsymbol{\varSigma}$ is a diagonal matrix containing the singular values $\sigma_{i}$ of $\mathbf{\mathbf{H}}$. The product $\mathbf{\mathbf{H}}\mathbf{H}^{H}$ can be written as
\begin{equation}
\mathbf{\mathbf{H}}\mathbf{H}^{H}=\mathbf{U}\boldsymbol{\varSigma}\mathbf{\boldsymbol{\varSigma}}^{H}\mathbf{U}^{H},
\end{equation}
where $\boldsymbol{\varSigma}\mathbf{\boldsymbol{\varSigma}}^{H}$ is a diagonal matrix, and therefore the inverse of $\left(\mathbf{\mathbf{H}}\mathbf{H}^{H}\right)^{-1}$ can be written as
\begin{equation}
\left(\mathbf{\mathbf{H}}\mathbf{H}^{H}\right)^{-1}=\mathbf{U}\left(\boldsymbol{\varSigma}\mathbf{\boldsymbol{\varSigma}}^{H}\right)^{-1}\mathbf{U}^{H}.
\end{equation}
Since $\boldsymbol{\varSigma}\mathbf{\boldsymbol{\varSigma}}^{H}$ is diagonal with entries $\mu_{i}^{2}$ , its inverse is
\begin{equation}
\left(\boldsymbol{\varSigma}\mathbf{\boldsymbol{\varSigma}}^{H}\right)^{-1}={\rm diag}\left(\frac{1}{\mu_{1}^{2}},\frac{1}{\mu_{2}^{2}},\dots\right).
\end{equation}
Now, we have
\begin{align}
{\rm Tr}\left(\left(\mathbf{\mathbf{H}}\mathbf{H}^{H}\right)^{-1}\right) & = {\rm Tr}\left(\mathbf{U}\left(\boldsymbol{\varSigma}\mathbf{\boldsymbol{\varSigma}}^{H}\right)^{-1}\mathbf{U}^{H}\right)\notag\\
 & ={\rm Tr}\left({\rm diag}\left(\frac{1}{\mu_{1}^{2}},\frac{1}{\mu_{2}^{2}},\dots\right)\right)\notag\\
 & = \sum_{i=1}^{r}\frac{1}{\mu_{i}^{2}},\label{eq:18-1}
\end{align}
where $r$ denotes the rank of the matrix $\mathbf{\mathbf{H}}$, and $\mu_{i}$ are the non-zero singular values of $\mathbf{\mathbf{H}}$. Thus, for the SVD-TIM algorithm, in each step of Algorithm \ref{alg:gstim}, we select one port that results in the largest decrease in the objective function by using (\ref{eq:18-1}), and stop when $n_{s}$ ports are selected. Using (\ref{eq:18-1}) instead of calculating the matrix inverse which has high complexity, it leads to reduction in the computational complexity. 

\begin{remark}
In radar-centric systems, the optimization problem can be reformulated to maximize the radar SINR while achieving a target secrecy rate as
\begin{eqnarray}
\underset{\boldsymbol{\Pi}_{n_{s}},\mathbf{W}}{\max} & \gamma_{b}\nonumber \\
\mbox{s.t.} & \sum_{k=1}^{K} R_{s}^{k}\geq r_{\rm th}\nonumber \\
 & \left[\boldsymbol{\Pi}_{n_{s}}\right]_{l}\in\left\{ \mathbf{z}_{1},\dots,\mathbf{z}_{N_{s}}\right\} \nonumber \\
 & \left[\boldsymbol{\Pi}_{n_{s}}\right]_{l}\neq\left[\boldsymbol{\Pi}_{n_{s}}\right]_{k}\nonumber \\
 & {\rm Tr}\left(\mathbf{W}\mathbf{W}^{H}\right)\leq P,\label{eq:42}
\end{eqnarray}
where $r_{\rm th}$ is the worst-case achieved sum secrecy rate. This problem can be solved by following similar steps of the JPPS scheme introduced in Algorithms \ref{alg:sca} and \ref{alg:jpps}.
\end{remark}

\section{Simulation Results\label{sec:Numerical-Results}}
In this section, we present numerical results to evaluate the performance of the proposed FAS-enabled secure ISAC framework. Following \cite{G10_new2024MIMO-FAS}, without loss of generality (WLOG), the FAS is placed at the $x\text{-}y$ plane and the receiving antennas are uniformly distributed along a straight line. Unless otherwise specified, the system parameters are set as follows: number of receive antennas is $N_{r}=10$, antenna area of $A_{s}=1\lambda$ with $f_{c}=2.4\textrm{ GHz}$, number of ports is $N_{s}=16$, number of activated ports is $n_{s}=6$, number of communication users is $K=4$, radar sensing SINR threshold is $\zeta=1$ and $m=2$. Moreover, the target is at a distance of $200\textrm{m}$ from the fluid antenna, and the spatial angles are generated randomly. The users are located at distances $2{\rm m}$, $15{\rm m}$, $25{\rm m}$, and $35{\rm m}$ from the center of the fluid antenna. For simplicity, equal noise variances are assumed at the users, $\sigma^{2}$. Hence, the transmit signal-to-noise ratio (SNR) is defined as ${\rm SNR}=\frac{P}{\sigma^{2}}$. 

\begin{figure}[]
\begin{centering}
\includegraphics[width=.9\columnwidth]{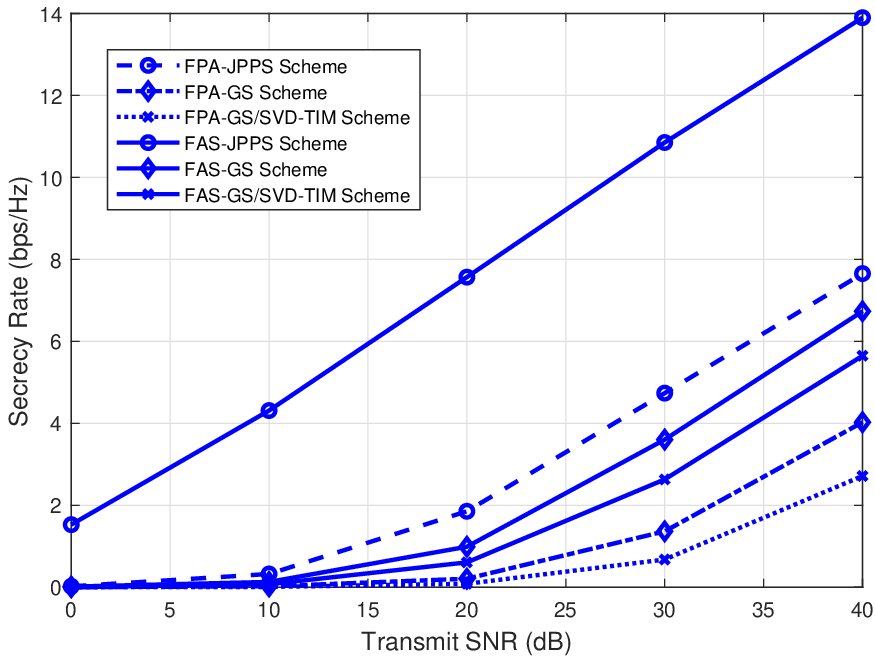}
\end{centering}
\caption{Secrecy rate versus the transmit SNR for FAS and FPA for the three schemes namely JPPS, GS and GS/SVD-TIM.}\label{fig:r1}
\end{figure}

Firstly, in Fig.~\ref{fig:r1}, we show the achievable sum secrecy rate versus transmit SNR for different antenna schemes, FAS and fixed-position antenna (FPA) of the proposed strategies. As expected, the secrecy rate increases with the transmit SNR due to the enhanced signal power. Moreover, it can be noted that FAS outperforms the fixed antenna configuration for all the considered scenarios. We also observe that JPPS achieves the highest secrecy rate across all SNR values, greatly outperforming all other counterparts. This gain is attributed to the enhanced spatial diversity and decorrelation afforded by the fluid antenna's dynamic repositioning capability, which is further exploited by the proposed utility-based optimization. For instance, at ${\rm SNR}=40~{\rm dB}$, JPPS achieves nearly $14~{\rm bps/Hz}$, whereas GS achieves around $7~{\rm bps/Hz}$. The performance gap between JPPS and the other heuristic methods (GS, and GS/SVD-TIM) highlights the importance of jointly optimizing both port selection and beamforming. 

\begin{figure}[]
\begin{centering}
\includegraphics[width=.9\columnwidth]{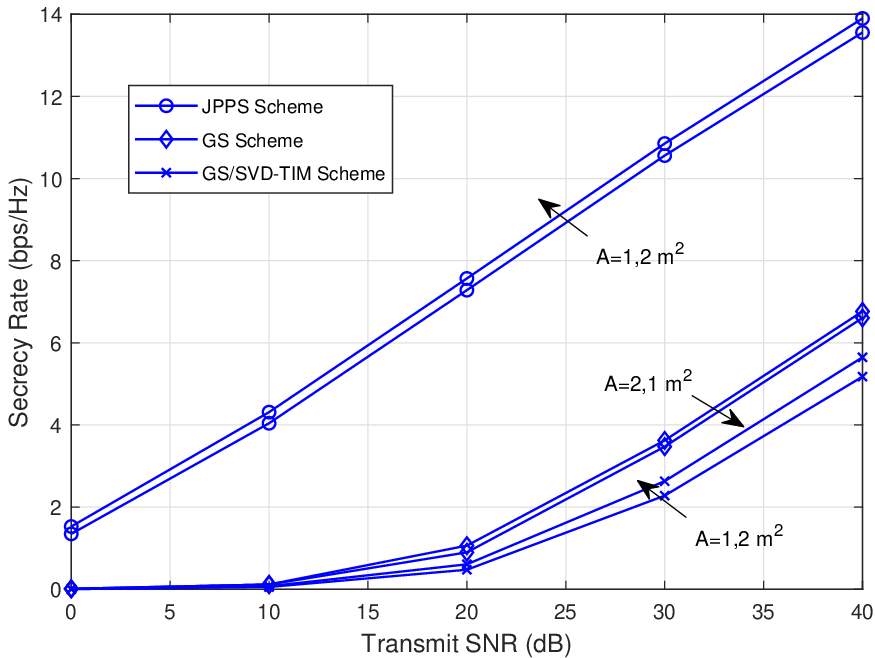}
\end{centering}
\caption{Secrecy rate versus the transmit SNR for different FAS area $A$ for the three schemes namely JPPS, GS and GS/SVD-TIM.}\label{fig:rVSsnr_dA}
\end{figure}

In Fig.~\ref{fig:rVSsnr_dA}, we illustrate the sum secrecy rate of the proposed algorithms versus the transmit SNR for different FAS surface areas, $A\in\{1,2\}$. It can be observed that increasing the FAS area improves the performance significantly. This is due to the fact that a larger area leads to increased spatial decorrelation between the ports, thereby providing greater DoF for both beamforming and port selection. Also, the use of the JPPS scheme enables effective joint optimization, ensuring that both communication and radar constraints are satisfied.

\begin{figure}[]
\begin{centering}
\includegraphics[width=.9\columnwidth]{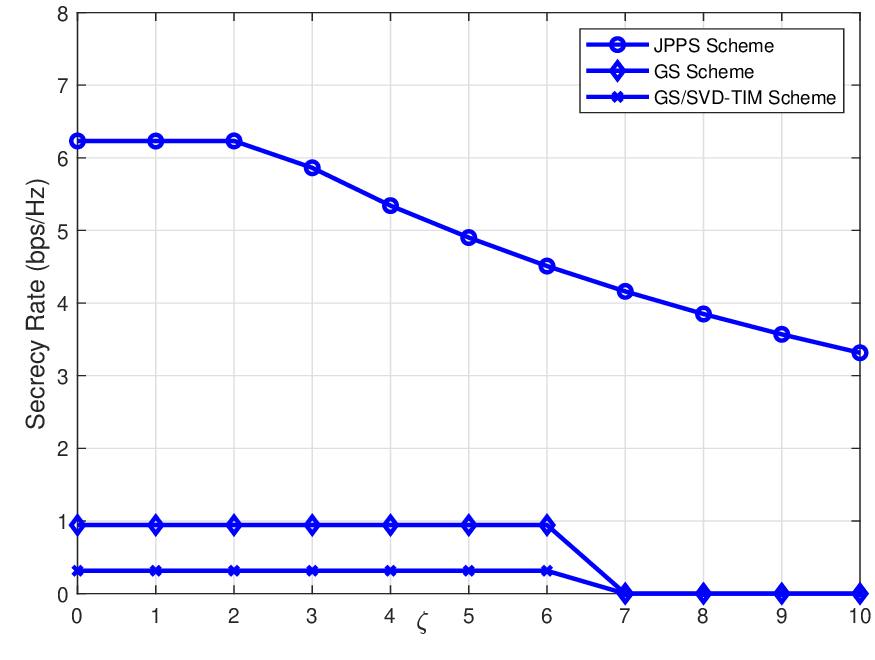}
\end{centering}
\caption{Secrecy rate versus the required radar sensing SINR for the three schemes namely JPPS, GS and GS/SVD-TIM.}\label{fig:rVSssnr}
\end{figure}

Fig.~\ref{fig:rVSssnr} presents the results showing how the sum secrecy rate changes as a function of the required radar SINR threshold, $\zeta$, when the target is located at a distance of $50{\rm m}$. This evaluation highlights the trade-off between radar sensing performance and communication secrecy. The results show that the secrecy rate decreases as the radar SINR constraint becomes more stringent. At $\zeta=0$, where radar sensing is not enforced, the JPPS scheme achieves a secrecy rate of about $6.2~{\rm bps/Hz}$. But as $\zeta$ increases, the radar constraint becomes dominant in the optimization problem, forcing more power to be directed toward the radar target, thus reducing the DoF available for secrecy enhancement. Thus, the secrecy rate drops steadily and reaches around $3~{\rm bps/Hz}$ at $\zeta\ge10$. In contrast, GS and GS/SVD-TIM exhibit minimal variation across different radar SINR levels. This indicates that these heuristic methods are not radar-aware and incapable of adjusting the transmission strategy to jointly satisfy both secrecy and sensing objectives. As a result, their secrecy rates remain consistently low (below $1~{\rm bps/Hz}$) regardless of the radar requirement. These results demonstrate the flexibility of the proposed JPPS framework, which can dynamically balance the dual objectives of secure communication and radar sensing, unlike other heuristics.

\begin{figure}
\begin{centering}
\includegraphics[width=.9\columnwidth]{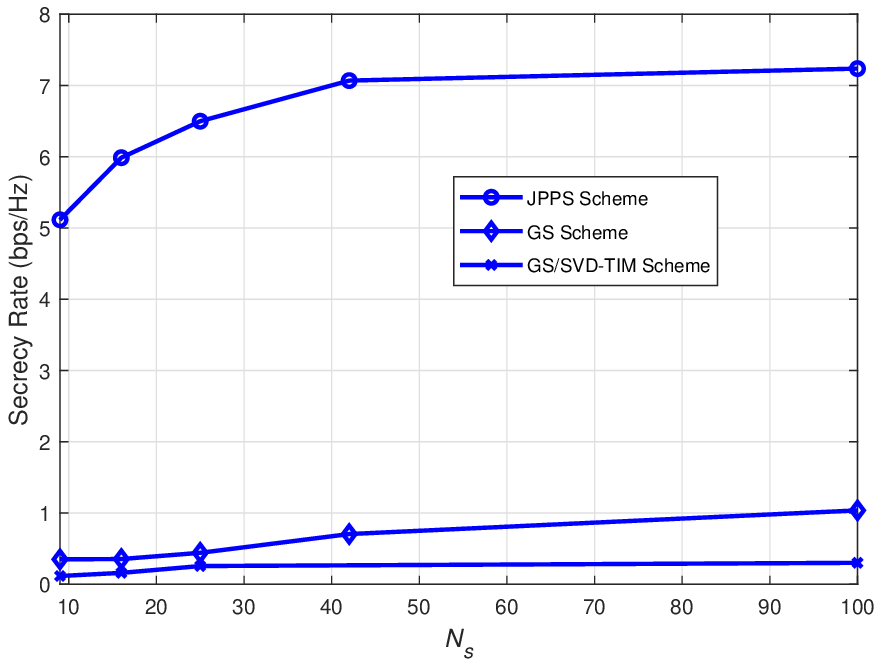}
\end{centering}
\caption{Secrecy rate versus the number of ports for the three schemes namely JPPS, GS and GS/SVD-TIM.}\label{fig:rVSNs}
\end{figure}

Fig.~\ref{fig:rVSNs} illustrates the impact of the total number of available FAS ports, $N_{s}$, on the achievable secrecy rate under the three schemes. This result emphasizes the scalability of the proposed solution and highlights how system flexibility benefits physical layer security. The proposed JPPS scheme exhibits a strong positive correlation between $N_{s}$ and the secrecy rate. As $N_{s}$ increases from $10$ to $100$, the secrecy rate improves substantially, reaching saturation at around $7.3~{\rm bps/Hz}$. This gain is due to the larger candidate pool of spatially distributed ports, which enables better port combinations for optimizing the beamforming structure and suppressing signal leakage to unintended users. The JPPS strategy exploits both the spatial diversity and fluidity of the antenna system, allowing it to adaptively select the most secure and effective ports. On the other hand, the GS and GS/SVD-TIM schemes demonstrate limited improvements. Their secrecy rates remain nearly constant or increase marginally with larger $N_{s}$, since they lack an intelligent joint design mechanism and rely on heuristic port selections. Even with more candidate ports, these methods are unable to leverage the full potential of the FAS architecture.

\begin{figure}
\begin{centering}
\includegraphics[width=.9\columnwidth]{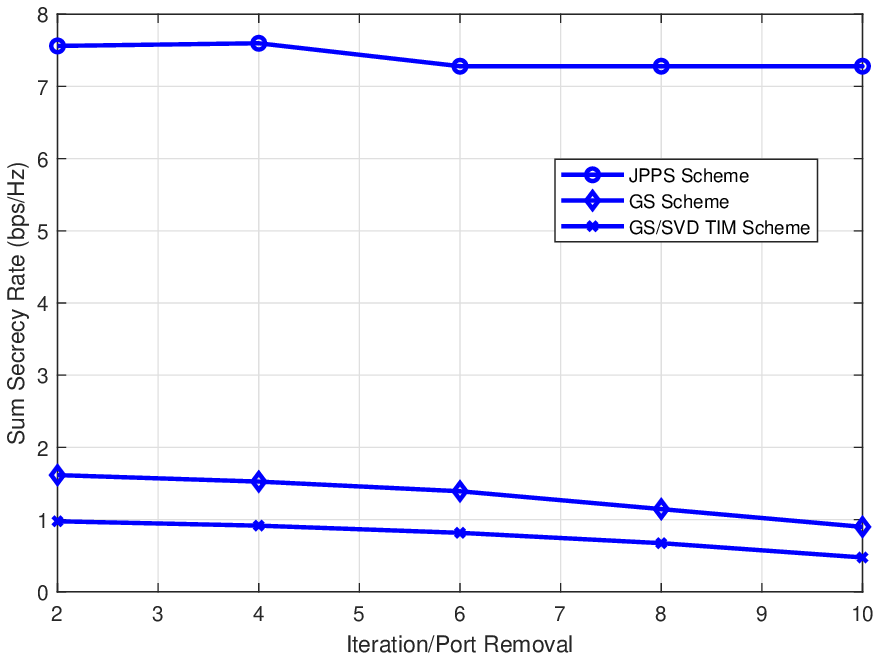}
\end{centering}
\caption{Secrecy rate versus the number of iterations for the three schemes namely JPPS, GS and GS/SVD-TIM.}\label{fig:rVSiteration}
\end{figure}

The convergence behavior of the proposed algorithms is now studied by the results in Fig.~\ref{fig:rVSiteration}, where the sum secrecy rate is plotted against the number of iterations. It is observed that the algorithms converge rapidly. Typically, JPPS converges within $6$ iterations while GS and GS-TIM converge within $10$ iterations. These results demonstrate the practical applicability and low computational burden of the proposed schemes. These results validate the efficiency of the proposed approaches.

\begin{figure}
\begin{centering}
\includegraphics[width=.9\columnwidth]{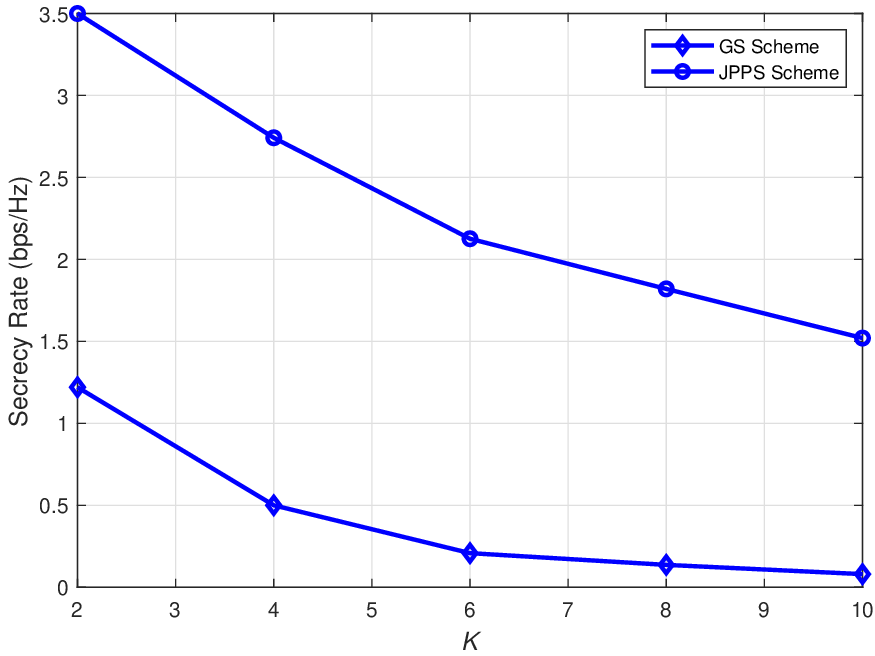}
\end{centering}
\caption{Secrecy rate versus the number of users $K$ for the JPPS and GS schemes when ${\rm SNR}=5~{\rm dB}$.}\label{fig:rVSK}
\end{figure}

Fig.~\ref{fig:rVSK} shows the sum secrecy rate versus the number of users, $K$, for the JPPS and GS algorithms when ${\rm SNR}=5~{\rm dB}$ and the users are uniformly distributed in a circle with a radius of $32{\rm m}$. As we can see, the sum secrecy rate of both schemes declines monotonically as $K$ increases. This is because the $K-1$ unintended users  act as passive eavesdroppers, which leads to an increase of the effective wiretap channel quality and decreases the secrecy rate for each legitimate link. Also, this performance degradation is attributed to a limited DoF. It can also be observed that JPPS consistently outperforms GS, demonstrating its superior ability to jointly optimize the port configuration and beamforming. These findings reinforce the advantage of the proposed joint optimization approach in enhancing physical layer security under a large user density.

\begin{figure}
\begin{centering}
\includegraphics[width=.9\columnwidth]{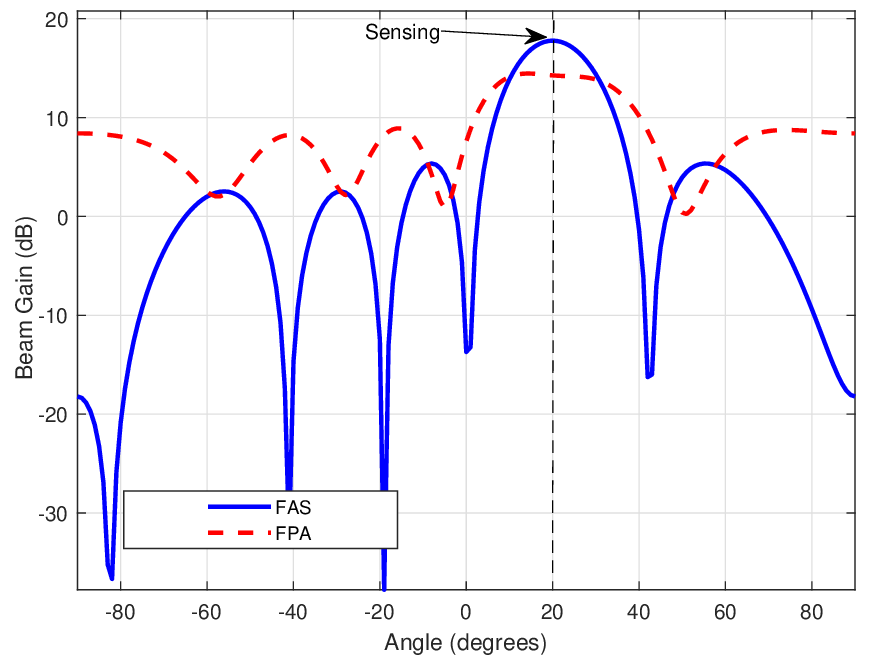}
\end{centering}
\caption{Beampatterns of FAS and FPA when $N_{s}=9$, $n_{s}=6$ and ${\rm SNR}=10~{\rm dB}$.}\label{fig:beamVSangle}
\end{figure}

Finally, the results in Fig.~\ref{fig:beamVSangle} compare the beampattern of the FAS and that of the FPA configuration in radar-centric scenarios, considered in (\ref{eq:42}). The beam gain is plotted across spatial angles, with a sensing direction marked at $20^\circ$. The results clearly show that the FAS-enabled system achieves sharper and more focused beamforming toward the target sensing direction, yielding a higher main-lobe gain. This demonstrates the superior adaptability of the FAS in aligning its radiation pattern toward the sensing target while maintaining secure multiuser communication. Such spatial agility highlights the significant advantage of using FAS in ISAC systems.

\section{Conclusion\label{sec:Conclusion}}
This paper presented a framework for enhancing physical layer security in ISAC systems through joint precoding and port selection in FAS. The proposed approach formulates a joint optimization problem to maximize the sum secrecy rate of multiple legitimate users while ensuring a minimum SINR radar constraint. A novel JPPS framework has been proposed to maximize the sum secrecy rate by jointly optimizing the transmit precoder and the FAS port selection using FP. Then ZF precoding was adopted and the secrecy rate maximization problem was reformulated and solved using greedy and trace inverse minimization schemes. Extensive simulations have demonstrated that FAS has better secrecy performance than fixed antenna configurations in all the considered scenarios. In addition, the proposed JPPS scheme outperforms benchmark approaches, including greedy and SVD-based heuristics, in terms of secrecy rate, radar sensing quality, and adaptability to the number of available ports and transmit power. 

\bibliographystyle{IEEEtran}

\begin{thebibliography}{10}
\providecommand{\url}[1]{#1}
\csname url@samestyle\endcsname
\providecommand{\newblock}{\relax}
\providecommand{\bibinfo}[2]{#2}
\providecommand{\BIBentrySTDinterwordspacing}{\spaceskip=0pt\relax}
\providecommand{\BIBentryALTinterwordstretchfactor}{4}
\providecommand{\BIBentryALTinterwordspacing}{\spaceskip=\fontdimen2\font plus
\BIBentryALTinterwordstretchfactor\fontdimen3\font minus
  \fontdimen4\font\relax}
\providecommand{\BIBforeignlanguage}[2]{{%
\expandafter\ifx\csname l@#1\endcsname\relax
\typeout{** WARNING: IEEEtran.bst: No hyphenation pattern has been}%
\typeout{** loaded for the language `#1'. Using the pattern for}%
\typeout{** the default language instead.}%
\else
\language=\csname l@#1\endcsname
\fi
#2}}
\providecommand{\BIBdecl}{\relax}
\BIBdecl
\bibitem{Wong2020Fluid}
K. K. Wong, K.-F. Tong, Y. Zhang, and Z. Zheng, ``Fluid antenna system for 6G: When Bruce Lee inspires wireless communications,'' {\em Elect. Lett.}, vol. 56, no. 24, pp. 1288--1290, Nov. 2020.
\bibitem{Wong2022Bruce}
K. K. Wong, K.-F. Tong, Y. Shen, Y. Chen, and Y. Zhang, ``{Bruce Lee}-inspired fluid antenna system: Six research topics and the potentials for {6G},'' {\em Frontiers Commun. Netw.}, vol. 3, Mar. 2022, Art. no. 853416.
\bibitem{New2025Tutorial}
W. K. New {\em et al.}, ``A tutorial on fluid antenna system for 6G networks: Encompassing communication theory, optimization methods and hardware designs,'' \emph{IEEE Commun. Surv. Tuts.}, vol. 24, no. 4, pp. 2325--2377, Aug. 2025.
\bibitem{Lu-2025}
W.-J. Lu {\em et al.}, ``Fluid antennas: Reshaping intrinsic properties for flexible radiation characteristics in intelligent wireless networks,'' {\em IEEE Commun. Mag.}, vol. 63, no. 5, pp. 40--45, May 2025.
\bibitem{FAS1}
K. K. Wong, A. Shojaeifard, K. F. Tong, and Y. Zhang, ``Performance limits of fluid antenna systems,'' {\em IEEE Commun. Lett.}, vol. 24, no. 11, pp. 2469--2472, Nov. 2020.
\bibitem{Fas2}
K. K. Wong, A. Shojaeifard, K. F. Tong, and Y. Zhang, ``Fluid antenna systems,'' {\em IEEE Trans. Wireless Commun.}, vol. 20, no. 3, pp. 1950--1962, Mar. 2021.
\bibitem{I24_shen2024design}
Y. Shen {\em et al.}, ``Design and implementation of mmWave surface wave enabled fluid antennas and experimental results for fluid antenna multiple access,'' {\em arXiv preprint}, \url{arXiv:2405.09663}, May 2024.
\bibitem{fas8}
J. Zhang {\em et al.}, ``A novel pixel-based reconfigurable antenna applied in fluid antenna systems with high switching speed,'' {\em IEEE Open J. Antennas \& Propag.}, vol.~6, no.~1, pp.~212--228, Feb. 2025.
\bibitem{Liu-2025arxiv}
B. Liu, K.-F. Tong, K. K. Wong, C.-B. Chae, and H. Wong, ``Programmable meta-fluid antenna for spatial multiplexing in fast fluctuating radio channels,'' {\em Optics Express}, vol. 33, no. 13, pp. 28898--28915, 2025
\bibitem{Tong-2025}
K.-F. Tong, B. Liu, and K. K. Wong, ``Designs and challenges in fluid antenna system hardware,'' {\em Electronics, Special Issue on Futuristic Antennas: Sustainable, Efficient, Reconfigurable, and Intelligent Design}, vol. 14, no. 7, pp. 1458, Apr. 2025.
\bibitem{Khammassi-2023}
M. Khammassi, A. Kammoun and M.-S. Alouini, ``A new analytical approximation of the fluid antenna system channel,'' {\em IEEE Trans. Wireless Commun.}, vol. 22, no. 12, pp. 8843--8858, Dec. 2023.
\bibitem{fas5}
J. D. Vega-S$\acute{\rm a}$nchez, L. Urquiza-Aguiar, M. C. P. Paredes and D. P. M. Osorio, ``A simple method for the performance analysis of fluid antenna systems under correlated Nakagami-$m$ fading,'' {\em IEEE Wireless Commun. Lett.}, vol. 13, no. 2, pp. 377--381, Feb. 2024.
\bibitem{Alvim-2023}
P. D. Alvim {\em et al.}, ``On the performance of fluid antennas systems under $\alpha$-$\mu$ fading channels,'' {\em IEEE Wireless Commun. Lett.}, vol. 13, no. 1, pp. 108--112, Jan. 2024.
\bibitem{G5_new2023SISO-FAS}
W. K. New, K. K. Wong, H. Xu, K. F. Tong and C. B. Chae, ``Fluid antenna system: New insights on outage probability and diversity gain,'' {\em IEEE Trans. Wireless Commun.}, vol. 23, no. 1, pp. 128--140, Jan. 2024.
\bibitem{H7_Espinosa2024Anew}
P. Ram\'{i}rez-Espinosa, D. Morales-Jimenez and K. K. Wong, ``A new spatial block-correlation model for fluid antenna systems,'' {\em IEEE Trans. Wireless Commun.}, vol.~23, no.~11, pp. 15829--15843, Nov. 2024.
\bibitem{G10_new2024MIMO-FAS}
W. K. New, K. K. Wong, H. Xu, K.-F. Tong and C.-B. Chae, ``An information-theoretic characterization of MIMO-FAS: Optimization, diversity-multiplexing tradeoff and $q$-outage capacity,'' {\em IEEE Trans. Wireless Commun.}, vol. 23, no. 6, pp. 5541--5556, Jun. 2024.
\bibitem{Psomas-dec2023}
C. Psomas, P. J. Smith, H. A. Suraweera and I. Krikidis, ``Continuous fluid antenna systems: Modeling and analysis,'' {\em IEEE Commun. Lett.}, vol. 27, no. 12, pp. 3370--3374, Dec. 2023.
\bibitem{Fas3}
Z.~Chai, K. K. Wong, K.-F. Tong, Y.~Chen, and Y.~Zhang, ``Port selection for fluid antenna systems,'' \emph{IEEE Commun. Lett.}, vol.~26, no.~5, pp. 1180--1184, May 2022.
\bibitem{G13_Efrem2024MIMO-FAS}
C. N. Efrem and I. Krikidis, ``Transmit and receive antenna port selection for channel capacity maximization in fluid-MIMO systems,'' {\em IEEE Wireless Commun. Lett.}, vol. 13, no. 11, pp. 3202--3206, Nov. 2024.
\bibitem{Chen-2025fmimo}
J.-C. Chen, T.-L. Cheng, K. K. Wong, and H. Shin, ``Improved joint transmit and receive port selection for capacity maximization in fluid-MIMO systems,'' {\em IEEE Wireless Commun. Lett.}, vol. 14, no. 6, pp. 1693--1697, Jun. 2025.
\bibitem{Feng-2025}
B. Feng, C. Feng, T. Q. S. Quek, and K. K. Wong, ``Deep unfolding neural networks for fluid antenna-enhanced vehicular communication,'' {\em IEEE Trans. Veh. Technol.}, vol. 74, no. 9, pp. 14793--14798, Sep. 2025.
\bibitem{Xu-fasup2025}
H. Xu {\em et al.}, ``Capacity maximization for FAS-assisted multiple access channels,'' {\em IEEE Trans. Commun.}, vol. 73, no. 7, pp. 4713--4731, Jul. 2025.
\bibitem{fas4}
K. K. Wong, K.-F. Tong, Y.~Chen, Y.~Zhang, and C.-B. Chae, ``Opportunistic fluid antenna multiple access,'' \emph{IEEE Trans. Wireless Commun.}, vol.~22, no.~11, pp. 7819--7833, Nov. 2023.
\bibitem{Waqar-2024}
N. Waqar {\em et al.}, ``Opportunistic fluid antenna multiple access via team-inspired reinforcement learning,'' {\em IEEE Trans. Wireless Commun.}, vol. 23, no. 9, pp. 12068--12083, Sep. 2024.
\bibitem{Hong2025FAS}
H. Hong {\em et al.}, ``FAS meets OFDM: Enabling wideband 5G NR,"  {\em IEEE Trans.  Commun.}, \url{doi: 10.1109/TCOMM.2025.3591751}, 2025.
\bibitem{Wang-2024oct}
C. Wang {\em et al.}, ``AI-empowered fluid antenna systems: Opportunities, challenges, and future directions,'' {\em IEEE Wireless Commun.}, vol. 31, no. 5, pp. 34--41, Oct. 2024. 
\bibitem{Wang-2025llm}
C. Wang, K. K. Wong, Z. Li, L. Jin, and C.-B. Chae, ``Large language model empowered design of fluid antenna systems: Challenges, frameworks, and case studies for 6G,'' to appear in {\em IEEE Wireless Commun.}, \url{arXiv:2506.14288}, 2025.
\bibitem{H4_wong2022FAMA}
K. K. Wong and K. F. Tong, ``Fluid antenna multiple access,'' \emph{IEEE Trans. Wireless Commun.}, vol.~21, no.~7, pp. 4801--4815, Jul. 2022.
\bibitem{H5_wong2023fast}
K. K. Wong, K. F. Tong, Y. Chen, and Y. Zhang, ``Fast fluid antenna multiple access enabling massive connectivity,'' {\em IEEE Commun. Lett.}, vol. 27, no. 2, pp. 711--715, Feb. 2023.
\bibitem{H6_wong2023sFAMA}
K. K. Wong, D. Morales-Jimenez, K. F. Tong, and C.-B. Chae, ``Slow fluid antenna multiple access,'' \emph{IEEE Trans. Commun.}, vol.~71, no.~5, pp. 2831--2846, May 2023.
\bibitem{H12_Wong2024cuma}
K. K. Wong, C. B. Chae, and K. F. Tong, ``Compact ultra massive antenna array: A simple open-loop massive connectivity scheme,'' {\em IEEE Trans. Wireless Commun.}, vol. 23, no. 6, pp. 6279--6294, Jun. 2024.
\bibitem{H10_hong2024coded}
H. Hong, K. K. Wong, K. F. Tong, H. Shin, and Y. Zhang, ``Coded fluid antenna multiple access over fast fading channels,'' \emph{IEEE Wireless Commun. Lett.}, vol.~14, no.~4, pp.~1249--1253, Apr. 2025.
\bibitem{H11_hong2025Downlink}
H. Hong {\em et al.}, ``Downlink OFDM-FAMA in 5G-NR systems,'' {\em IEEE Trans. Wireless Commun.}, \url{doi: 10.1109/TWC.2025.3577771}, Jun. 2025.
\bibitem{Waqar-tfama2025}
N. Waqar, K. K. Wong, C.-B. Chae, and R. Murch, ``Turbocharging fluid antenna multiple access,'' {\em IEEE Trans. Wireless Commun.}, \url{doi: 10.1109/TWC.2025.3607824}, 2025.
\bibitem{Han-2025}
T. Han, Y. Zhu, K. K. Wong, G. Zheng, and H. Shin, ``Cell-free fluid antenna multiple access networks,'' {\em IEEE Trans. Wireless Commun.}, vol. 24, no. 9, pp. 7237--7251, Sep. 2025. 
\bibitem{xu2024channel}
H. Xu {\em et al.}, ``Channel estimation for FAS-assisted multiuser mmWave systems,'' {\em IEEE Commun. Lett.}, vol. 28, no. 3, pp. 632--636, Mar. 2024.
\bibitem{fas7}
B. Xu, Y. Chen, Q. Cui, X. Tao and K. K. Wong, ``Sparse Bayesian learning-based channel estimation for fluid antenna systems,'' {\em IEEE Wireless Commun. Lett.}, vol. 14, no. 2, pp. 325--329, Feb. 2025.
\bibitem{fas9}
Z. Zhang, J. Zhu, L. Dai and R. W. Heath, ``Successive Bayesian reconstructor for channel estimation in fluid antenna systems,'' {\em IEEE Trans. Wireless Commun.}, vol. 24, no. 3, pp. 1992--2006, Mar. 2025.
\bibitem{Ref1}
A.~Liu {\em et al.}, ``A survey on fundamental limits of integrated sensing and communication,'' \emph{IEEE Commun. Surv. \& Tut.}, vol.~24, no.~2, pp. 994--1034, Secondquarter 2022.
\bibitem{Ref2}
D.~Wen {\em et al.}, ``A survey on integrated sensing, communication, and computation,'' \emph{IEEE Commun. Surv. \& Tut.}, \url{DOI: 10.1109/COMST.2024.3521498}, 2025.
\bibitem{Generalized}
L.~Chen, Z.~Wang, Y.~Du, Y.~Chen, and F.~Richard~Yu, ``Generalized transceiver beamforming for {DFRC} with {MIMO} radar and MU-MIMO communication,'' \emph{IEEE J. Select. Areas Commun.}, vol. 40, no. 6, pp. 1795--1808, Jun. 2022.
\bibitem{joint}
X.~Liu {\em et al.}, ``Joint transmit beamforming for multiuser {MIMO} communications and {MIMO} radar,'' \emph{IEEE Trans. Sig. Proc.}, vol.~68, pp. 3929--3944, Jun. 2020.
\bibitem{Hybrid}
Z.~Cheng, Z.~He, and B.~Liao, ``Hybrid beamforming for multi-carrier dual-function radar-communication system,'' \emph{IEEE Trans.  Cog. Commun. \& Netw.}, vol.~7, no.~3, pp. 1002--1015, Sep. 2021.
\bibitem{Spectral}
B.~Tang and J.~Li, ``Spectrally constrained {MIMO} radar waveform design based on mutual information,'' \emph{IEEE Trans. Sig. Proc.}, vol.~67, no.~3, pp. 821--834, Feb. 2019.
\bibitem{OFDM}
J.~Johnston, L.~Venturino, E.~Grossi, M.~Lops, and X.~Wang, ``{MIMO} {OFDM} dual-function radar-communication under error rate and beampattern constraints,'' \emph{IEEE J. Select. Areas Commun.}, vol. 40, no. 6, pp. 1951--1964, Jun. 2022.
\bibitem{Rev1ref2}
B.~Chang, W.~Tang, X.~Yan, X.~Tong, and Z.~Chen, ``Integrated scheduling of sensing, communication, and control for mmWave/THz communications in cellular connected UAV networks,'' \emph{IEEE J. Select. Areas Commun.}, vol.~40, no.~7, pp. 2103--2113, Jul. 2022.
\bibitem{meisac}
A.~Salem, K.~Meng, C.~Masouros, F.~Liu, and D.~Lopez-Perez, ``Rethinking dense cells for integrated sensing and communications: A stochastic geometric view,'' \emph{IEEE Open J. Commun. Soc.}, vol.~5, pp. 2226--2239, 2024.
\bibitem{survy}
F.~Liu, C.~Masouros, A.~P. Petropulu, H.~Griffiths, and L.~Hanzo, ``Joint radar and communication design: Applications, state-of-the-art, and the road ahead,'' \emph{IEEE Trans. Commun.}, vol.~68, no.~6, pp. 3834--3862, Jun. 2020.
\bibitem{Ref3}
Z.~Yang {\em et al.}, ``Secure precoding optimization for NOMA-aided integrated sensing and communication,'' \emph{IEEE Trans. Commun.}, vol.~70, no.~12, pp. 8370--8382, Dec. 2022.
\bibitem{Ref4}
W.~Sun, S.~Sun, X.~Su, and R.~Liu, ``Security-ensured integrated sensing and communication (ISAC) systems enabled by phase-coupled intelligent omni-surfaces (IOS),'' \emph{IEEE Trans. Wireless Commun.}, vol.~23, no.~4, pp. 3480--3492, Apr. 2024.
\bibitem{Ref5}
J.~Chu, R.~Liu, M.~Li, Y.~Liu, and Q.~Liu, ``Joint secure transmit beamforming designs for integrated sensing and communication systems,'' \emph{IEEE Trans. Veh. Technol.}, vol.~72, no.~4, pp. 4778--4791, Apr. 2023.
\bibitem{Ref6}
Y.~Cao, L.~Duan, and R.~Zhang, ``Sensing for secure communication in ISAC: Protocol design and beamforming optimization,'' \emph{IEEE Trans. Wireless Commun.}, vol.~24, no.~2, pp. 1207--1220, Feb. 2025.

\bibitem{Meng-2025}
K. Meng, C. Masouros, K. K. Wong, A. P. Petropulu, and L. Hanzo, ``Integrated sensing and communication meets smart propagation engineering: Opportunities and challenges,'' {\em IEEE Netw.}, vol. 39, no. 2, pp. 278--285, Mar. 2025.
\bibitem{H17_wang2024fluid}
C. Wang {\em et al.}, ``Fluid antenna system liberating multiuser MIMO for ISAC via deep reinforcement learning,'' {\em IEEE Trans. Wireless Commun.}, vol.~23, no.~9, pp. 10879--10894, Sep. 2024.
\bibitem{Zhou-2024fisac}
L. Zhou, J. Yao, M. Jin, T. Wu, and K. K. Wong, ``Fluid antenna-assisted ISAC systems,'' {\em IEEE Wireless Commun. Lett.}, vol. 13, no. 12, pp. 3533--3537, Dec. 2024. 
\bibitem{Ye-2025}
Y. Ye {\em et al.}, ``SCNR maximization for MIMO ISAC assisted by fluid antenna system,'' {\em IEEE Trans. Veh. Technol.}, vol. 74, no. 8, pp. 13272--13277, Aug. 2025.
\bibitem{H16_Zou2024shifting}
J. Zou {\em et al.}, ``Shifting the ISAC trade-off with fluid antenna systems,'' \emph{IEEE Wireless Commun. Lett.}, vol.~13, no.~12, pp.~3479--3483, Dec. 2024.
\bibitem{Tang-2025fisac}
B. Tang {\em et al.}, ``Full-duplex FAS-assisted base station for ISAC,'' {\em IEEE Trans. Wireless Commun.}, \url{doi: 10.1109/TWC.2025.3600362}, 2025.
\bibitem{Lou-2025}
X. Lou, W. Xia, Y. Zhu, K. K. Wong, and C.-B. Chae, ``Multi-target beamforming optimization for fluid antenna-enabled multi-static ISAC,'' {\em IEEE Trans. Cog. Commun. \& Netw.}, \url{doi: 10.1109/TCCN.2025.3612765}, 2025.
\bibitem{H15_ghadi2024perf}
F. Rostami Ghadi, K. K. Wong, F. J. L\'{o}pez-Mart\'{i}nez, H. Shin, and L. Hanzo, ``Performance analysis of FAS-aided NOMA-ISAC: A backscattering scenario,'' {\em IEEE Internet Things J.}, \url{doi: 10.1109/JIOT.2025.3612478}, 2025.

\bibitem{mvdr1}
Y.~Gu and A.~Leshem, ``Robust adaptive beamforming based on interference covariance matrix reconstruction and steering vector estimation,'' \emph{IEEE Trans. Sig. Proc.}, vol.~60, no.~7, pp. 3881--3885, Jul. 2012.
\end{thebibliography}

\end{document}